\documentclass{ar-1col-S2O}
\usepackage[numbers]{natbib}
\usepackage{url}
\usepackage{amsfonts}
\usepackage{amsmath}
\newcommand{\comments}[1]{}   %%%%%%%%%%%%%%% comments

\jname{Xxxx. Xxx. Xxx. Xxx.}
\jvol{AA}
\jyear{2023}
\doi{10.1146/((please add article doi))}

\begin{document}

% Page header
\markboth{L. F. Cugliandolo}{Non-equilibrium dynamics}

% Title
\title{Recent Applications of Dynamical Mean-Field Methods}

%Authors, affiliations address.
\author{Leticia F. Cugliandolo$^{1,2}$
\affil{$^1$Laboratoire de Physique Th\'eorique et Hautes Energies, Sorbonne Universit\'e, UMR 7589,
4 Place Jussieu, 75252 Paris Cedex 05, France; \\
email: leticia@lpthe.jussieu.fr}
\affil{$^2$Institut Universitaire de France,\\
1 rue Descartes, 75231 Paris Cedex 05, France}
}

%Abstract
\begin{abstract}
Rich out of equilibrium collective dynamics of strongly interacting large assemblies 
emerge in many areas of science. Some intriguing and not fully
understood examples are the glassy arrest in atomic, molecular or colloidal systems,  
flocking in natural or artificial active matter,  and the organization and subsistence of ecosystems. 
The learning process, and ensuing amazing performance, of deep neural networks bears resemblance with 
some of the before-mentioned examples. Quantum 
mechanical extensions are also of interest. In exact or approximate manner the  evolution
of these systems  can be expressed in terms of a dynamical mean-field theory which not only captures many of their 
peculiar effects but  also has predictive power. This short review presents a summary of recent 
developments of this approach with emphasis on applications on the examples mentioned above.
\end{abstract}

%Keywords, etc.
\maketitle

%Table of Contents
\tableofcontents

% Heading 1
\section{INTRODUCTION}

In condensed matter or statistical physics focus is set on
macroscopic systems in thermal equilibrium. 
%Microcanonical, canonical or grand canonical ensembles are used, depending on the conditions one is
%interested in. 
The relaxation of a tiny perturbation away from
equilibrium is also sometimes described in textbooks and undergraduate
courses. However, the vast majority of physical systems evolve far from equilibrium.

Take a classical macroscopic physical system in conventional equilibrium.
Any change in the  properties of the
environment  in a canonical setting, or in the system's Hamiltonian itself in a microcanonical
one, will take it away  from equilibrium.  The procedure of rapidly (ideally instantaneously) tuning a
parameter is called a {\it  quench}.  
%Right after a  quench the system's configuration is not one of equilibrium
%under the new conditions and the system subsequently evolves in an out of
%equilibrium fashion.
The post-quench relaxation towards the new equilibrium (if
possible) could be fast or very slow, and even need a time scaling with the system size.
Well-known cases include systems quenched to a critical point~\cite{Hohenberg-Halperin} or
%which evolve through critical dynamics, or 
across a second order phase transition~\cite{Bray94,Onuki02,Puri09,HenkelPleimling10}.
% which later undergo domain growth 
Less well-understood cases, as those with 
competing interactions that behave as glasses have exceedingly long relaxations as well~\cite{BeBi11}. The post-quench dynamics of closed systems 
%and, in particular, energy conserving ones 
are also of great interest at present, boosted by cold atom experiments performed in almost perfect isolation~\cite{Bloch08}.

Out of equilibrium situations can also be established by {\it external drives}. 
In the context of macroscopic physics,  a 
traditional example is Rayleigh-B\'enard convection 
induced in a layer of fluid by two parallel confining plates maintained at different temperatures.
Another appealing instance is the one of 
powders stuck in static metastable states unless tapping, vibration or shear 
make them slowly evolve towards more compact configurations~\cite{Duran99}.
Active matter, with manifold physical, biological and artificial realizations
is another example, in which energy is injected at the
microscopic constituent level~\cite{Bechinger16}. 

The out of equilibrium evolution of systems with many interacting components is not restricted to physics.
Celebrated areas in which tools learnt and developed to deal with 
complex physical systems are currently being applied with success are theoretical ecology~\cite{Maritan16}, neuroscience~\cite{Bahri19}, computer science~\cite{Krzakala13}, 
and econophysics~\cite{Naumann-Woleske23}, to name just a few.

The problems discussed in the previous paragraphs pertain to the classical
World. In other  cases of practical interest, quantum
fluctuations play an important r\^ole. The main focus of this review  will be  the description of classical 
non-equilibrium dynamics, but we will comment on applications of highlighted notions to quantum ones
as well.

Characterizing the macroscopic behavior of far from equilibrium systems is challenging. One cannot 
rely on any ergodic hypothesis to compute statistical averages and, in contrast,  full time dependencies have to be elucidated. Much 
inspiration has been gained from the study of simplified {\it microscopic} models that do capture
 important features of coarsening, glassy and weakly driven systems. These approximate mathematical 
representations are often {\it bona fide} ones for problems in other, and at least as interesting, 
areas of science. We present a set of models that have been key to the construction of quite a general picture 
for the non-equilibrium dynamics of complex systems with slow dynamics in Sec.~\ref{sec:models}.
The review goes on with the recollection of  methods and ideas developed in the 
study of their relaxational dynamics
in Secs.~\ref{sec:DMFT}, \ref{sec:rugged-free-energies}, and \ref{sec:relaxation-dynamics}. 
We then focus on recent applications and discoveries.
In Sec.~\ref{sec:hamilton-dynamics} we describe the peculiarities of their Hamiltonian evolution. 
Section~\ref{sec:driven-dynamics} is devoted to the discussion of cases explicitly maintained far from equilibrium by the 
action of non-potential forces. Applications to active matter and ecology pertain to this class.
We also comment of recent progress in the analysis of stochastic learning processes of neural networks.
In relaxational cases, rough free-energy landscapes play a crucial role in 
trapping the dynamics. In the driven ones the complexity translates to the random forces.
Finally, we briefly discuss quantum extensions and relations to recent work in high-energy physics in Sec.~\ref{sec:quantum}.
We close with conclusions and some open problems  in Sec.~\ref{sec:Conclusions}.
Several lecture notes and review articles give a more technical presentation of the formalism discussed here.
The reader interested in learning the mathematical derivations can consult~Refs.~\cite{LesHouches,Cavagna09,Montanari22,Manacorda}.

\section{MODELS}
\label{sec:models}

Consider the set of coupled differential equations
\begin{eqnarray}
m \frac{d^2x_i(t)}{dt^2} +  \eta \frac{dx_i(t)}{dt} = g_i(x_i) + f_i(x_1,\dots,x_N) + h_i(t) + \xi_i(t)
\; , 
\label{eq:defining-eqs}
\end{eqnarray}
ruling the time dependence of a typically large number of real-valued variables $x_i(t)$, 
$i=1,\dots, N$.  In their simplest interpretation, these variables represent the Cartesian coordinates of 
the position vector $\vec x = (x_1, \dots, x_N)$ of a point-like particle with mass $m$
 moving in an $N$-dimensional space. The space of configurations is continuous and differentiable.
% Let us discuss the elements in Equations~\ref{eq:defining-eqs} one by one.

The inertial term in the left-hand-side is important to model Hamiltonian dynamics as well as to establish quantum extensions. 
In the Langevin case it gives rise to 
{\it under-damped} dynamics, while when negligible the dynamics get  {\it over-damped}. 

The last term in the right-hand-side is a white Gaussian thermal noise with 
%vanishing average and correlations
\begin{eqnarray}
\langle \xi_i(t)\rangle = 0 
\; , 
\qquad\qquad
\langle \xi_i(t) \xi_j(t') \rangle = 2\eta k_BT \, \delta_{ij} \delta(t-t')
\; , 
\end{eqnarray}
which, together with the friction terms $\eta \dot x_i$ 
%(last one in the left-hand-side), 
model the coupling of  $x_i$ to an equilibrated bath at temperature $T$. The 
coupling strength is controlled by the friction coefficient $\eta$. 
The balance between the noise correlation and friction kernel
%, both delta functions for white noise, 
ensures that the environment is in thermal equilibrium. Henceforth, noise averages 
are indicated with angular brackets $\langle \dots \rangle$.

The coupling to colored baths which are not necessarily in equilibrium could also be used. 
For still Gaussian noises, the friction term and noise-noise correlations  generalize to 
\begin{equation}
\int_0^t dt' \; \Gamma_R(t,t') \, \frac{dx_i(t')}{dt'}
\; , 
\qquad\qquad\qquad
\langle \xi_i(t) \xi_j(t')\rangle = \delta_{ij} \, \Gamma_C(t,t')
\; . 
\end{equation}
The kernels satisfy  $\Gamma_R(t,t') \propto \theta(t-t')$ (causality) and $\Gamma_C(t,t')=\Gamma_C(t',t)$
 (symmetry) and represent the linear response and correlation of the bath. 
 %For an 
 In equilibrium both kernels are time-translationally invariant, 
 meaning that they only depend on time-difference $t-t'$. A fluctuation-dissipation-theorem (FDT) 
 holds: $ \Gamma_C (t-t') = k_BT \, [\Gamma_R(t-t') + \Gamma_R(t'-t)] $ or equivalently $\Gamma_R(t-t') = \beta 
 \theta(t-t') \Gamma_C (t-t')$. Equilibrium white noise  is recovered with $\Gamma_C (t-t') = k_BT \, \eta/\tau
 \,  e^{-|t-t'|/\tau}$  and $\Gamma_R(t-t') = \eta/\tau  \, e^{-|t-t'|/\tau} \theta(t-t')$
in the limit in which the characteristic time vanishes, $\tau\to 0$, implying $\Gamma_C (t-t') \to 2\eta k_BT \, \delta(t-t')$
and $\Gamma_R (t-t') \to 2\eta  \delta(t-t') \theta(t-t')$.
Some active matter models  use this last $\Gamma_R(t-t')$ 
with $\Gamma_C (t-t') = 2 \eta k_BT \, \delta(t-t') + \Gamma_{\rm act}(t-t')$, 
where $\Gamma_{\rm act}(t-t')$ is interpreted as being due to particle self-propulsion 
and violates FDT by construction.

%In the following 
We will mostly use equilibrium white noise. The coupling to 
the environment can then 
be on by keeping $\eta\neq 0$ or switched off by setting  $\eta=0$. The noise can be eliminated by 
putting $T$  to zero, rendering the {\it stochastic Langevin} Equations~\ref{eq:defining-eqs} {\it deterministic}.

The time-dependent functions $h_i(t) $ are external 
forces. They can be, for example, sinusoidal,  $h_i(t) = h
\sin(\omega t)$, and relevant to discuss periodically driven systems, or just constant as in an infinitesimal 
perturbation applied to measure a linear response.

The term $g_i(x_i)$  is a local force which could restrict the
values of $x_i$ for $g_i(x_i) = -V_i'(x_i)$ 
with $V_i(x_i) = u(x^2_i-a^2_i)^2$ and 
favor  $x_i = \pm a_i$, or it could impose a global 
spherical constraint
%These equations can be complemented by a global or $N$ local constraints 
%on the variables $x_i$, 
%such as a spherical one or a positivity ones:
\begin{equation}
\sum_{i=1}^N x_i^2 = N  \qquad \Rightarrow  \qquad 
g_i(x_i) = z(t) x_i
%\;\; {\rm (spherical)}
%\qquad\mbox{or} 
%\qquad
%x_i \geq 0 \;\;\; \forall \; i \;\; \;\; {\rm (semi-positive)} 
\; ,
\label{eq;constraints}
\end{equation}
 (on average) with the  Lagrange multiplier $z(t)$  to be  determined self-consistently. 
In this case, the particle is forced to move
 % in Equations~\ref{eq;constraints} then 
 on the sphere of radius $N^{1/2}$ and $\langle \vec x \cdot \frac{d\vec x}{dt} \rangle = 0$.
% (again, on average).
 % (spherical) or in the quadrant in which all coordinates are positive or vanish (semi-positive).
%respectively.

\begin{marginnote}[]
\entry{Fully-connected}{Each variable interacts with all others.}
\end{marginnote}
The internal forces $f_i(x_1,\dots,x_N)$ are generically {\it multi-body}. 
We will concentrate on {\it fully-connected} models, such that each variable interacts with all other ones, 
and the forces are independent of any distance.
%have no spatial structure, that is, 
Moreover, we will consider {\it disordered} $f_i$, which depend on quenched random parameters drawn from 
a time-independent probability distribution. 
\begin{marginnote}[]
\entry{Quenched disorder}{Time-independent parameters drawn from a probability distribution.}
\end{marginnote}

%Concrete examples will be specified below.

Whenever the forces are {\it conservative}, $f_i = -\partial V/\partial x_i$, we focus on
a Gaussian distributed potential energy $V$ with 
 $[V(\vec x)] =0$ and
%\begin{equation}
$[V(\vec x) V(\vec y)] = N \; \widetilde V\left(|\vec x -\vec y|^2/N \right)$.
%\; . 
%\label{eq:potential-correlations0}
%\end{equation}
\begin{marginnote}[]
\entry{Conservative forces}{Those that derive from a potential energy.}
\end{marginnote}
Henceforth, $[\dots ]$ denote average over disorder.
The scaling with $N$ ensures an interesting thermodynamic limit. 
For notational simplicity and also to ease the understanding, we collect 
all $x_i$ in the $N$ dimensional vector $\vec x = (x_1, \dots, x_N)$.

The $p$-spin models, originally proposed for Ising variables~\cite{MePaVi}
but later generalized to spherical  ones~\cite{CrSo92}, 
\begin{equation}
V(\vec x) = - \sum_{i_1 < \dots < i_p} \!\!\! J_{i_1\dots i_p} x_{i_1} \dots  x_{i_p}
\; , 
\label{eq:p-spin}
\end{equation}
with a fixed integer, $p\geq 2$, 
belong to this category.  
The couplings $J_{i_1\dots i_p}$,  symmetric under the permutation of any pair of indices,   
are {\it i.i.d.} quenched random variables 
Gaussian distributed with $[J_{i_1\dots i_p}] =0$,  $[J_{i_1\dots i_p}^2] = p! J^2/(2 N^{p-1})$,
and  uncorrelated for different sets of indices. The 
potential energy correlation then reduces to
\begin{equation}
[V(\vec x) V(\vec y)] = N \; \overline  V(\vec x \cdot \vec y/N)
% = N \left(\frac{\vec x \cdot \vec y}{N} \right)^p
\qquad
\mbox{with}
\qquad \overline V(q) = (J^2 /2) \, q^p
\; . 
\label{eq:potential-correlations}
\end{equation}
At zero temperature ($T=0$), vanishing external forces 
($h_i=0$), and in the absence of the inertial term ($\eta/m \gg 1$), 
%and in the potential case ($f_i = - \partial V/\partial x_i$), 
the potential dynamics boil down to 
a {\it gradient descent}. 
Switching off the coupling to the bath ($\eta=0$), the  sum of kinetic and potential energies is {\it conserved}.
 For special $V$ there can be other conserved quantities and even as many
as the number of degrees of freedom, making the dynamics {\it integrable} 
 (see Sec.~\ref{sec:hamilton-dynamics}).
%(see Sec.~\ref{subsec:Neumann}).

%We will focus on cases in which 
%they derive from a potential energy $V(x_1, \dots,x_N)$, and take the form $f_i = - \partial V/\partial x_i$, or they do not, and there is no $V$ such that $f_i = - \partial V/\partial x_i

{\it Non-reciprocal forces} 
%do not derive from a potential, $f_i \neq -\partial V/\partial x_i$, 
do not satisfy Newton's action-reaction principle. 
Simple extensions of the model defined in Equation~\ref{eq:p-spin} provide 
examples of this kind; {\it e.g.},
\begin{marginnote}[]
\entry{Non-reciprocal forces}{those that violate the action-reaction principle.}
\end{marginnote}
%. For instance, one can achieve a force 
%exerted by $j$ on $i$ that is not equal to the force exerted by $i$ on $j$ with  the form 
\begin{equation}
f_i(\vec x) = 
( \sum_{(i<) j_2 < \dots < j_p } 
+ 
\sum_{j_2 (< i <) j_3 < \dots < j_p } 
+
\dots + 
\sum_{j_2<  \dots < j_p (< i)} ) \; 
J_{ij_2 \dots j_p} x_{j_2} \dots x_{j_p}
\; , 
\end{equation}
where each constrained multi-sum runs over $p-1$ indices, 
if the exchanges $J_{ij_2\dots j_p}$ are
symmetric under permutations of the $j_2, \dots, j_p$ indices but they are not under permutations of the 
index $i$ with any of the $j$s~\cite{Sompolinsky88,CuKuLePe97,Berthier00,Fyodorov}  (see Sec.~\ref{sec:driven-dynamics}).  
They have zero mean, are uncorrelated for different indices, and have variance $p!/(2N^{p-1})$.
Then, $[f_i(\vec x)]=0$ and
\begin{eqnarray}
[f_i(\vec x) f_j(\vec y)] &=& \delta_{ij} \; N \, \overline V'\left(\frac{\vec x \cdot \vec y}{N}\right) 
+  
x_j  y_i \; \overline V''\left(\frac{\vec x \cdot \vec y}{N}\right)
+
(\gamma-1) \, x_j  y_i \; \overline V''\left(\frac{\vec x \cdot \vec y}{N}\right)
\; .
\label{eq:force-correlation}
\end{eqnarray}
The conservative case is recovered for $\gamma=1$.

The dynamics are usually
studied on average over initial conditions $\{ x_i(0), \, \dot x_i(0) \}$ 
 sampled  from distributions which reflect specific situations, like thermal equilibrium
at a prescribed temperature, or knowledge of the forces $f_i$.  Preparing initial 
conditions in canonical equilibrium at $\beta_0$ and evolving them with Langevin 
dynamics at $\beta$ is one  way to perform a {\it quench}.
Notation-wise, when there is thermal 
noise, we include the average over initial conditions in $\langle \dots \rangle$
while for deterministic dynamics we denote it $\langle \dots \rangle_{i.c.}$.
\begin{marginnote}[]
\entry{Quench}{A sudden change in a control parameter that sets the system out of equilibrium.}
\end{marginnote}

This set of equations has sufficient  ingredients to encompass, at least at some level of 
approximation, a large number of many-body physical phenomena, including the ones listed as examples in the 
Introduction. Furthermore, they also mimic problems in
other areas of science, such as neural networks~\cite{Bahri19}, optimization and 
computer science~\cite{Krzakala13}, evolution and game theory~\cite{Hofbauer98} or even markets~\cite{Naumann-Woleske23}. 
Because of the complicated nature of the $f_i$, 
these problems are dubbed {\it complex systems}. 
In interesting cases the interactions are strong and perturbative methods fail badly; self-consistent 
approaches, as the ones discussed in Sec.~\ref{sec:DMFT}, are better suited to deal with them.
%Finally, we wish to comment on 
Quantum extensions 
(Sec.~\ref{sec:quantum}) 
%A quantum Hamiltonian for a 
%single particle embedded in an $N$ dimensional space and under the effect of a random potential with the
%characteristics given above can be readily written,  
%and a spherical constraint can also be formulated~\cite{CuLo98,CuLo99,Cugliandolo02,Busiello07,CuLoNe19,Thomson20,Anous21,Bera22,Winer22,Correale23}. 
%The particle can be considered in isolation or coupled to a bath, modeled, for example with the 
%Feynman-Vernon procedure~\cite{Feynman63} leading to an influence functional. These models 
have been considered 
in the statistical physics literature and have also become fashionable among high-energy 
physicists
%~\cite{Anous21,Bera22} 
since the Sachdev-Ye-Kitaev (SYK) model, a fermionic analog of the classical $p$ body model, is 
a toy model of holography.

\section{DYNAMICAL MEAN-FIELD METHODS}
\label{sec:DMFT}

There are two alternative ways to exploit the fully-connected character of the models at hand, and write manageable equations for their evolution. One way derives deterministic
coupled integro-differential equations for correlation and linear response functions of Schwinger-Dyson 
kind. The other represents the full dynamics with an effective non-Markov Langevin equation for a single degree of freedom. 
We schematically present them here. 

\subsection{Schwinger-Dyson equations}

In the $N\to\infty$ limit  the dynamics of the spherical models are fully characterized by the 
two-time global correlation and linear response. The equations ruling their evolution can be derived
with  the  Martin-Siggia-Rose-Jansen-DeDominicis (MSRJD) functional  formalism~\cite{Martin73,Janssen76,DeDominicis76,Arnoulx23b} 
\begin{marginnote}[]
\entry{MSRJD}{
The functional integral generating function of Langevin processes.}
\end{marginnote}
or with perturbative methods
combined with self-consistent diagram re-summation~\cite{Bouchaud96}.
A common choice is to weight  the initial conditions 
with the Gibbs-Boltzmann measure 
% = {\beta_0}^{-1}$ 
 \begin{equation}
P(\{x_i(0), \dot x_i(0) \}) = Z^{-1}(\beta_0) \, {\rm e}^{-\beta_0 H_{0}(\{x_i(0),\dot x_i(0))\}}
 \end{equation}
at temperature $T_0$ with $H_0$ the  pre-quench Hamiltonian. For $T_0^{-1} = \beta_0\to 0$ they 
 are completely random while for $\beta_0>0$ they have knowledge of the Hamiltonian $H_0$.
 (Hereafter we  measure $T$ in units of $k_B$.)
 In order to average over them, when $H_0$ depends on quenched random parameters, one resorts to the replica 
 method~\cite{MePaVi}. 
 \begin{marginnote}[]
\entry{Replica method}{
A technique to compute the disorder averaged free-energy.}
\end{marginnote}
 %The initial  variables are replicated, $x_i(0) \mapsto x_i^a(0)$ with $a=1,\dots, n$.
  When the 
 quench concerns a change in temperature but not in the Hamiltonian, or it consists in a global rescaling of the 
 random coupling strengths, the replica structure of the initial conditions, which depends on $\beta_0$ and $H_0$, is conserved in time.
 Choosing, for simplicity, $T_0\geq T^0_s$, the static critical temperature of the model $H_0$, 
 the replica structure is rather simple ({\it symmetric}) and the dynamics can then be fully encoded in the  two-time 
correlation and linear response:
\begin{eqnarray}
%\qquad\qquad
 C(t_1,t_2) =  N^{-1} \sum_{i=1}^N \; [\langle x_i(t_1) x_i(t_2)\rangle] \; ,
%\\
%\qquad\qquad
% C(t_1,0) &=& N^{-1} \sum_{i=1}^N \; [\langle x_i(t_1) x_i(0)\rangle] \; ,
%\\
\qquad\quad
R(t_1,t_2) =  \left. N^{-1}  \sum_{i=1}^N \; [\langle \frac{\delta {x}^{(h)}_{i}(t_1)}{\delta h_i(t_2)} \rangle]
\right|_{h=0} \!\!\!\!\!\!\!\!\!\! \; ,
\;\;\;
\end{eqnarray}
for $t_1, \ t_2 > 0$, where $\vec x$ represents any of the replicas, as they are all equivalent.
The infinitesimal perturbation $\vec h$ is coupled linearly to the $x_i$ according to $H\mapsto H - \sum_{i=1}^N h_i x_i$ at time $t_2$
and the upper-script ${(h)}$ indicates that the configuration is measured under the field $\vec h$.
%The square brackets denote  the average over quenched disorder. 
%The angular brackets indicate the average over
%thermal noise if the system is coupled to an environment, and over the initial conditions. 
The boundary values of $C$ and $R$ at $t_1=t_2$ depend on whether the inertia term is present and whether the system is open or closed. 
Naming $z(t_1)$ is the Lagrange multiplier which enforces the spherical constraint $C(t_1,t_1)=1$, using an homogeneous external field $h(t)$, 
and calling
\begin{equation}
G_0^{-1}(t_1) = m\partial_{t_{1}}^{2}+\eta\partial_{t_{1}}+z(t_{1})
\; , 
\label{eq:G0}
\end{equation} 
the Schwinger-Dyson equations read~\cite{SompolinskyZippelius81,CuKu93,BaBuMe96,CuLoNe17,CuLoNe19}
\begin{marginnote}[]
\entry{Schwinger-Dyson equations}{
Closed integro-differential equations for correlations and linear responses.}
\end{marginnote}
\begin{eqnarray}
&& G_0^{-1}(t_1)
%\left(m\partial_{t_{1}}^{2}+\gamma\partial_{t_{1}}+z(t_{1})\right)
R(t_{1},t_{2}) =
\delta(t_1-t_2)
+
\int_{t_{2}}^{t_{1}}dt\; 
%\frac{J^2 p(p-1)}{2}R(t_{1},t)C^{p-2}(t_{1},t)
\Sigma(t_1,t) R(t,t_{2})
\; ,
\label{eq:dyn-eqs-R}
\\
&& G_0^{-1}(t_1)
%\left(m\partial_{t_{1}}^{2}+\gamma\partial_{t_{1}}+z(t_{1})\right)
C(t_{1},t_{2})=
\int_{0}^{t_{1}}dt \; 
%\frac{J^2 p(p-1)}{2}R(t_{1},t)C^{p-2}(t_{1},t)
\Sigma(t_1,t) C(t,t_{2})
+ \int_{0}^{t_{2}}dt \;  D(t_1,t)  R(t_{2},t) + \; \eta T R(t_2, t_1) 
%\frac{J^2 p}{2} C^{p-1}(t_{1},t)
\nonumber\\
 &  &
\qquad\qquad\qquad\qquad 
+ \, h(t_1) \int_0^{t_2} dt \; h(t) R(t_2, t)
+ 
%%%% this factor is for changing the couplings too ! \frac{J_0}{J} 
\beta_0
D(t_1,0)
%\frac{J^2 p}{2} C^{p-1}(t_{1},0)
C(t_{2},0) \; ,
%%%+\frac{x J^2 p}{2T}\sum_a (C_a^{p-1}(t_{1},0))^{p-1} C_a(t_{2},0)
\;
%%%%
\label{eq:dyn-eqs-C}
\end{eqnarray}
\begin{eqnarray}
&& z(t_{1}) =
\int_{0}^{t_{1}}dt \; [\Sigma(t_1,t) C(t_1,t) + D(t_1,t) R(t_{1},t)  ] 
-m\partial_{t_{1}}^{2}C(t_{1},t_{2})\vert_{t_{2}\rightarrow t_{1}^{-}}
%%%%+ \eta T \, R(t_1,t_2) \vert_{t_{2}\rightarrow t_{1}^{-}}
\nonumber\\
&& 
\qquad\quad
+ \eta T R(t_2 \to t^-_1, t_1)
+ h(t_1) \int_0^{t_1} \! dt \, h(t) R(t_1, t)
%\frac{JJ_0 p}{2T_0}C^{p}(t_{1},0)
%\frac{J^2 p^{2}}{2}C^{p-1}(t_{1},t)
+
%%%% this factor is for changing the couplings too ! \frac{J_0}{J} 
\beta_0 D(t_1,0)C(t_1,0) 
\label{eq:dyn-eqs-z}
\; .
%%%%% \frac{x J^2 p}{2T'} \sum_a (C_a(t_{1},0))^p
\end{eqnarray}
%%%where we set, for simplicity, $k_B=1$.
%We will specify them when needed. 
%with $\gamma =1$ the conservative case. 
The self-energy and vertex kernels, $\Sigma$ and $D$, depend on the correlation of the forces $f_i$, 
\begin{eqnarray}
\Sigma= \overline V''\!(C) R 
\qquad{\rm and}\qquad
D = \overline V'(C) \qquad \Rightarrow \qquad\Sigma =  D' R
\label{eq:Sigma-D}
\end{eqnarray}
 for the potential cases in Equation~\ref{eq:force-correlation}.
The terms proportional to $J_0/T_0$ impose the initial conditions.
% In particular, they vanish for completely random ones, for which $T_0\to\infty$. 
Importantly enough, out of equilibrium the 
linear response and correlation function are not necessarily related to one another and 
the equilibrium fluctuation dissipation theorem
\begin{equation}
T \, R(t_1, t_2) = \frac{\partial}{\partial t_2} C(t_1,t_2) \, \qquad \mbox{for} \;\; t_1 \geq t_2
\qquad\qquad
{\mbox{(FDT)}}
\end{equation} 
does not need to hold. Each two-time function should be determined  independently.
\begin{marginnote}[]
\entry{FDT}{
A model independent relation between linear response and correlation functions valid in equilibrium.}
\end{marginnote}

In quantum models one can proceed similarly, using the closed-time-path or Schwinger-Keldysh path integral~\cite{Sc61,Ke64}
formalism and derive the corresponding Schwinger-Dyson equations. This can be 
done even for special choices of initial states, as the canonical equilibrium ones~\cite{Stefanucci,CuLoNe19,Chakraborty19}.

\subsection{Single variable effective representation}

An alternative method allows one to analyze
the dynamics of interacting degrees of freedom in terms of a
{\it self-consistent one-body}  problem. 
The single variable effective Langevin equation 
can be derived from the MSRJD dynamic generating 
functional averaged over the quenched randomness~\cite{SompolinskyZippelius81,CuKu93,Agoritsas18}
\begin{marginnote}[]
\entry{Single variable effective description}{
The self-consistent one-body problem.}
\end{marginnote}
\begin{equation}
%m \frac{d^2x}{dt^2}  + \eta \frac{dx}{dt}  = g(x(t)) 
G_0^{-1}(t_1) x(t_1) + \int_0^{t_1} dt' \; \tilde\Sigma(t_1,t') x(t')
= \beta_0 \tilde D(t_1,0) x(0)  + h(t_1) + \zeta(t_1) + \xi(t_1)
\; . 
\label{eq:DMFT}
\end{equation}
The representative variable is $x$ and $G_0^{-1}$ is the differential 
operator defined in Equation~\ref{eq:G0}. 
There are two 
uncorrelated noises: the original zero-average white noise $\xi$, and
a new effective noise $\zeta$, also with zero mean and 
\begin{equation}
\langle \zeta(t_1) \zeta(t_2) \rangle = \tilde D(t_1,t_2) 
\; . 
\label{eq:self-cons-noise}
\end{equation}
The vertex $\tilde D$ acts as the colored noise correlation and the self-energy $\tilde \Sigma$ as the time-derivative of 
a retarded friction in the non-Markovian Langevin Equation~\ref{eq:DMFT}. In the potential case 
they take the same functional forms as in 
Equation~\ref{eq:Sigma-D}
%The kernels $\tilde\Sigma$ and $\tilde D$ take 
% for the $p$-spin models. 
%We repeat them here for the potential case
%\begin{equation}
%\tilde \Sigma = J^2 \, \frac{p(p-1)}{2} \, \tilde C^{p-2} \tilde R 
%\; , 
%\qquad\qquad
%\tilde D = J^2 \, \frac{p}{2}  \, \tilde C^{p-1}  
%\; ,
%\end{equation}
with 
\begin{equation}
\tilde  C(t_1,t_2) = 
%\frac{1}{N} \sum_{i=1} 
x(t_1) x(t_2) \; , 
\qquad\qquad\quad
\tilde R(t_1,t_2)  =  
% \frac{1}{N} \sum_{i=1} 
\left. \frac{\delta x(t_1)}{\delta h(t_2)} \right|_{h=0}
\!\!\!\! 
\end{equation}
and they collect the effects of the other degrees of freedom on the selected one.  
Note that $\tilde C$ and $\tilde R$ are not averaged over the thermal noises but in the  $N\to\infty$ limit the fluctuations 
should be suppressed~\cite{BenArous06}, and $\langle \tilde C\rangle  \to C$ and $\langle \tilde R \rangle \to R$, 
where the angular brackets represent an average over both $\xi$ and $\zeta$. 
The global spherical constraint transforms in 
$\langle x^2(t)\rangle = 1$. The set of equations has to be solved self-consistently; the difficulty lies in imposing the 
condition~\ref{eq:self-cons-noise}. 

The single variable equation can also be derived with an extension of the static {\it cavity} method
to the time-dependent problem~\cite{MePaVi,Agoritsas18,Roy19}. 
This technique is particularly useful since it allows one
to treat problems with non-linear  $g_i(x_i)$, and find that the 
linear term $z(t_1) x(t_1)$ is replaced by $g'_i(x_i)$
%term in Equation~\ref{eq:defining-eqs} are non-linear 
and, also, ``noise multiplicative'' cases, as the ecological models  
of Sec.~\ref{subsec:ecology}. The analysis in~\cite{Agoritsas18}
 gives a very detailed explanation of how to 
sample $x(0)$ and $\zeta(0)$ so as to comply with equilibrium conditions at 
$\beta_0$. 
\begin{marginnote}[]
\entry{Cavity method}{
a generalization of the Bethe—Peierls iterative technique used to obtain the static properties of models defined on 
tree-like graphs.
}
\end{marginnote}

%Details can be found in~\cite{MePaVi,Agoritsas18,Roy19}.

This approach bears similarity to the Dynamical Mean-Field Theory~\cite{Georges96} which has been 
so successfully applied to condensed matter systems, though mostly in equilibrium.

\section{RUGGED FREE-ENERGY LANDSCAPES}
\label{sec:rugged-free-energies}

In potential cases, the Thouless-Anderson-Palmer (TAP)
free-energy landscape~\cite{MePaVi} that generalizes the Ginzburg-Landau one to cases with  $N$ 
order parameters can be constructed and studied in great detail, see Refs.~\cite{Ros23,Auffinger23} for recent reviews. 
In order to set the stage for the discussion of the relaxation of conservative systems we
briefly recall their main features.
\begin{marginnote}[]
\entry{TAP free-energy landscape}{An extension of the Ginzburg-Landau free-energy to cases with $N$ local order parameters.}
\end{marginnote}

\subsection{The spherical $p=2$ case}

The spherically constrained random harmonic potential  is still complex enough  to allow for an equilibrium phase
transition~\cite{KoThJo76} but the structure of the free-energy landscape is rather simple. 
%the intermediate temperature range collapses. 
There are only two equilibrium states at $T<T_s$ related by $x_i \mapsto -x_i$ for all $i$, 
akin to a conventional second order phase transition with spontaneous symmetry breaking.
 The ground states are $\vec x= \pm \vec v^{\; \rm max}= \pm \vec v_{\mu=N}$, the 
eigenvector associated to the largest (ordered so that $\mu=N$) eigenvalue of the GOE coupling matrix ${\mathbb J}$.  
The  equilibrium states at $T<T_s$ just dress these configurations with thermal fluctuations, 
see the sketch in Fig.~\ref{fig:sketch-states}(b). The configurations 
closely aligned with the other eigenvectors of  ${\mathbb J}$ belong to metastable states. They 
are ordered according to their stability and energy, with the maximum associated to the eigenvector with the lowest eigenvalue ($\mu=1$).
Under a constant and uniform field $\vec h$ the number of stationary points of the potential energy is reduced to just two
with one of the zero-field ground states being selected by the field. There is no finite temperature phase 
transition under a field.  

Under spontaneous symmetry breaking,  $\langle x_{\mu=N}\rangle_{\rm eq}
\equiv \langle \vec x \cdot \vec v_{N}\rangle_{\rm eq} = (q_{\rm EA}N)^{1/2}$ and one mode is macroscopically occupied
until $T_s$ where  $q_{\rm EA}$ vanishes. This special alignment resembles Bose-Einstein condensation
and for this reason one calls the equilibrium low-temperature configurations {\it condensed} and the 
high temperature ones {\it extended}. 

Within replica theory, 
this model is {\it replica symmetric} (RS)
since the Parisi overlap matrix,  ${\mathbb Q} = (Q_{ab} = N^{-1} \sum_i [\langle x_i^a x_i^b \rangle])$, 
is filled with identical values $Q_{a\neq b} = q_{\rm EA}$. The parameter $q_{\rm EA}$ quantifies  the ``size'' of the 
equilibrium states.

 \subsection{The Sherrington-Kirkpatrick model}
 
In  the Sherrington-Kirkpatrick model of spins glasses, $p=2$ and
%is the most celebrated representative of this class. In this model 
$x_i = \pm 1$, there is a phase transition at $T_s$.
The replica structure at $T<T_s$ needs a {\it full replica symmetry breaking} (FRSB)
and this entails a hierarchical ({\it ultrametric}) organization of equilibrium states as the one in 
Fig.~\ref{fig:free-energy-sketches}(c). In a few words, there are equilibrium states of all 
kinds, in the sense that the overlap between two configurations in equilibrium can take any value $0\leq q < q_{\rm EA}$.
Barriers between them and the metastable states scale sub-linearly with $N$~\cite{MePaVi}.

%LATER
%The Gardner phase of the bimodal $p\geq 3$ model is another example of this kind. In these cases there is 
%no threshold level in the sense that there are no stable metastable states above the equilibrium ones.

\subsection{The spherical $p\geq 3$ potentials}

\begin{figure}[h!]
%\hspace{1cm}
%$(a)$ \hspace{3cm} $(b)$ \hspace{3cm} $(c)$ 
%\\
\centerline{
\hspace{-2.5cm}
\includegraphics[scale=0.2]{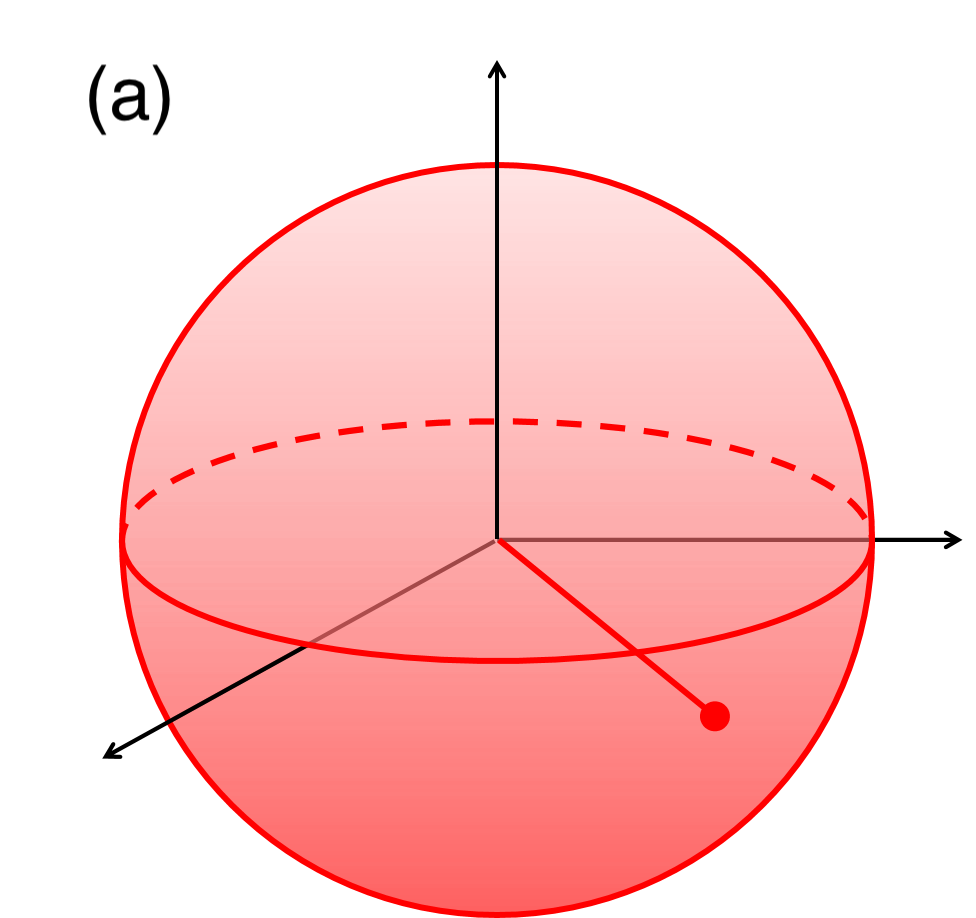}
\hspace{-8.25cm}
\includegraphics[scale=0.2]{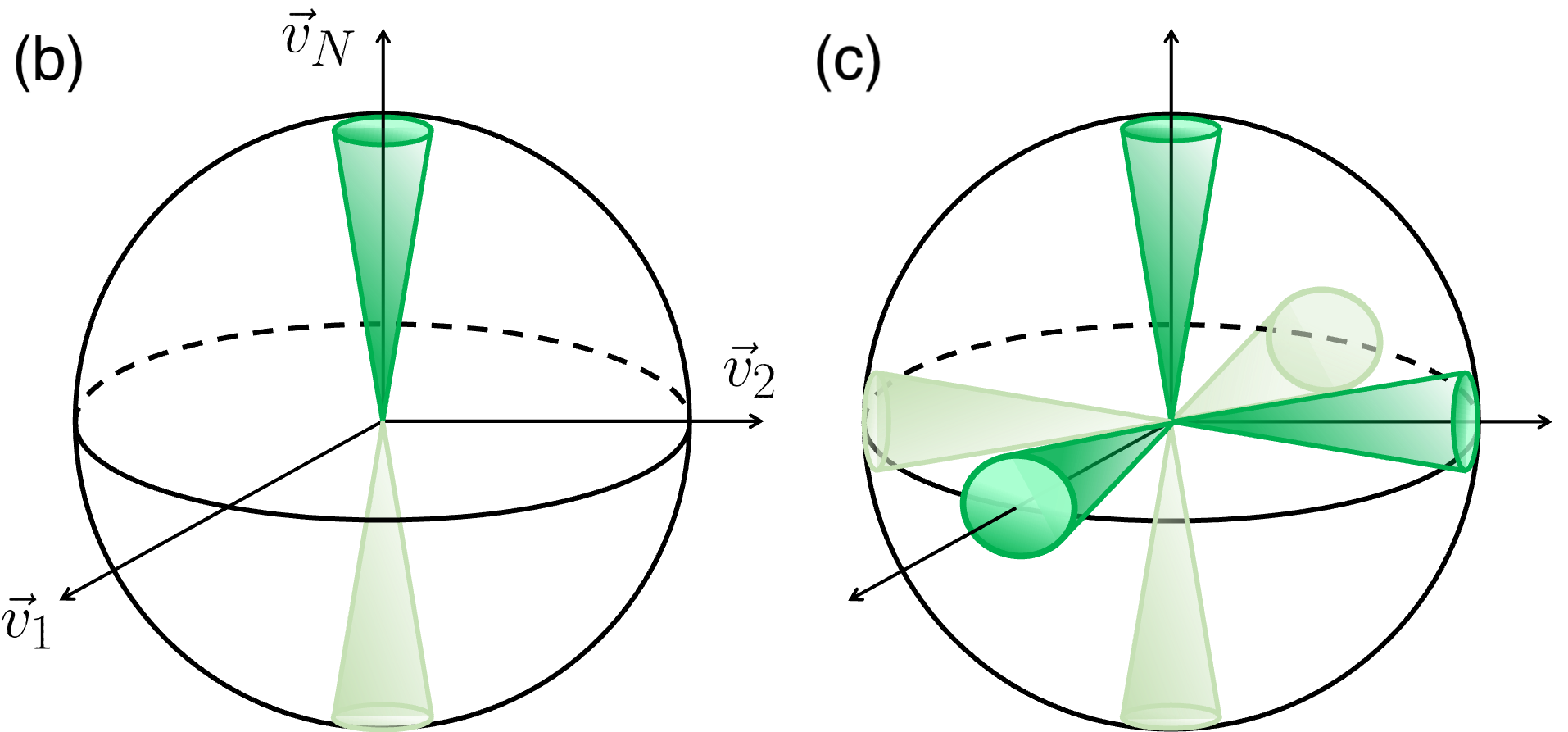}
}
\vspace{0.5cm}
\caption{The $N$ dimensional sphere.
(a) A high temperature (extended) situation with no metastable states, both in the $p=2$ and $p=3$ models.
(b) In the $p=2$,  the axes are the eigenvectors $\vec v_1, \dots \vec v_N$
of the interaction matrix ${\mathbb J}$. Only $\vec v_N$ the eigenvector with largest 
eigenvalue (and its reversed) are equilibrium 
states and are highlighted in green. There are no metastable states. 
(c) In the $p\geq 3$ case there are orthogonal metastable states,  some of 
them shown in the sketch.
}
\label{fig:sketch-states}
\end{figure}

\begin{figure}[b!]
%\hspace{1.5cm} (a) \hspace{3cm} (b) \hspace{3cm} (c)\\
\centerline{
%\raisebox{0.2cm}{
\includegraphics[scale=0.16]{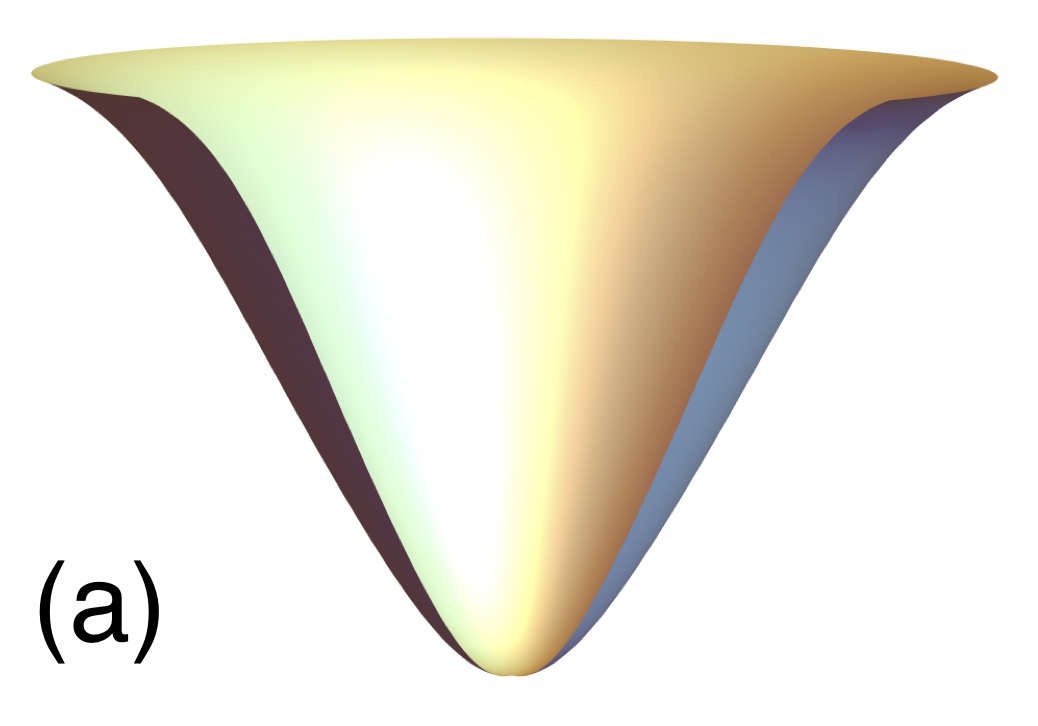}
%}
\hspace{-9.5cm}
\includegraphics[scale=0.15]{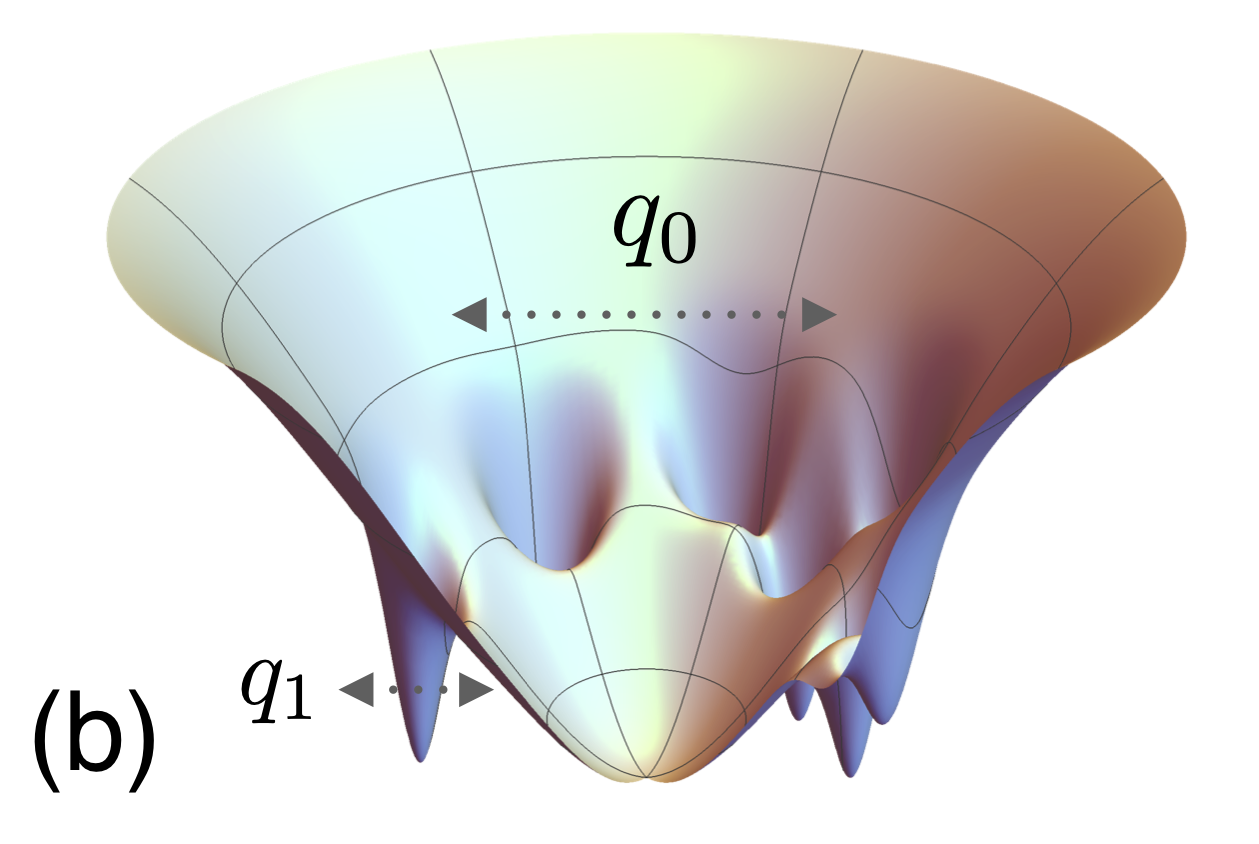}
\hspace{-9.5cm}
\raisebox{-0.1cm}{
\includegraphics[scale=0.22]{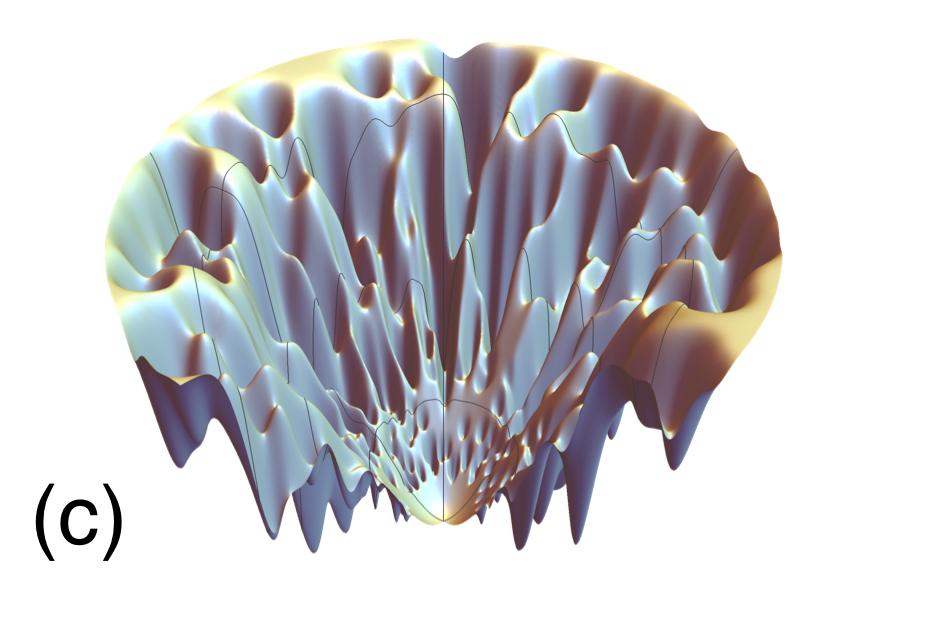}
}
}
\vspace{0.3cm}
\caption{Sketches of the free-energy density landscape close to equilibrium. (a) A single minimum, typical of high temperatures where
all effects of the interactions are washed out by the strong thermal fluctuations. 
(b) Multiple minima of the $p\geq 3$ model at temperatures below $T_s$. The parameters
$q_0$ and $q_1=q_{\rm EA}$ indicate the overlap between configurations of different equilibrium states and configurations 
within the same one, respectively. (c) The hierarchical organization of equilibrium states with relatively low 
barriers between them as realized in the Sherrington-Kirkpatrick model and Gardner phases. Higher lying metastable states 
and flat regions as the threshold of the $p\geq 3$ model are not represented in these sketches but can be found in 
Fig.~\ref{fig:free-energy-landscape-p3}.}
\label{fig:free-energy-sketches}
\end{figure}

The spherical $p\geq 3$ models, defined by the potential energy in Equation~\ref{eq:p-spin},
belong to a different {\it universality class} from all points of view (static, landscape, and dynamic).
Their TAP free-energy landscape is complex~\cite{BrMo80,KuPaVi92,CrSo95,Monasson95,CaGiPa98,CaGaGi99}
with manifold consequences. 
\begin{enumerate}
%\item[(i)]
\item
For $T > T_d$ there is a single global minimum, with vanishing local order parameters, $\langle x_i\rangle = 0$
(paramagnetic/liquid).
%, which dominates the equilibrium properties.
Any two typical configurations in equilibrium at these temperatures are orthogonal. 
(Sketch in Fig.~\ref{fig:sketch-states}(a) and Fig.~\ref{fig:free-energy-sketches}(a).)
%\item[(ii)]
\begin{marginnote}[]
\entry{Complexity}{The logarithm of the number of stationary points of the TAP free-energy landscape.}
\end{marginnote}
\item
In the interval $T_s < T < T_d$ there is ergodicity breaking with an exponentially large number of (metastable) states, 
that is, a finite {\it complexity}, defined as the  logarithm of the number of such states (also called
{\it configurational entropy}).
Metastable states are sets of configurations separated by free-energy barriers whose height scale as  $N$. 
The averages of the $x_i$
over all configurations belonging to the same state are the $N$ order parameters,  $\langle x_i \rangle \neq 0$. 
The stable states are in one-to-one 
correspondence with local minima of the potential energy.
They can be followed in temperature until a spinodal where they disappear
without crossing, merging nor dividing~\cite{KuPaVi92}.
%which may lie above the ground state. 
%The energy of the corresponding minimum can then be used to label the state.
Two configurations from the same state have finite overlap $q_1$ while 
two configurations from different states are orthogonal and have vanishing overlap $q_0=0$
(sketches in Fig.~\ref{fig:sketch-states}(c) and \ref{fig:free-energy-sketches}(b)).
In spite of the existence of so many metastable states in this range of temperatures,  
the equilibrium properties are the simple continuation of  the high temperature ones. 
Their behavior at $T\stackrel{>}{\sim} T_d$ is like the one of super-cooled liquids~\cite{KiTh87b}. 
%\item[(iii)]
\item
For $T<T_s $ the lowest lying states dominate the equilibrium measure, and have zero 
complexity ({\it entropy crisis}). 
This drives the static glass transition, 
which is thermodynamically second order (the energy is continuous and there is no latent heat), 
but the global order parameter is discontinuous and jumps  to a finite value as in a first order transition.   
This entropy vanishing transition, is the analog of the 
empirically-defined {\it Kauzmann transition}, where a crossing of the configurational entropy of 
the crystal and the glass is envisioned to occur in real materials~\cite{Debenedetti96}.  The thermodynamic properties below $T_s$ are specific to the glass.  
\begin{marginnote}[]
\entry{Entropy crisis}{Vanishing complexity.}
\end{marginnote}
\end{enumerate}
These features are at the basis of the  {\it Random First Order Transition} (RFOT) 
scenario for fragile glasses~\cite{KiTh15,Wolynes23}.
\begin{marginnote}[]
\entry{RFOT}{
A scenario for the glassy arrest.}
\end{marginnote}

The stability of the stationary points of the free-energy density is determined by the 
spectrum of eigenvalues of the Hessian, concretely, the left-side edge
of a semi-circle law:
\begin{enumerate}
\item
 If it is larger than zero, all eigenvalues are positive and 
it is a stable minimum.
 \item
When it is negative, the saddle point is unstable with negative eigenvalues.
\item
If it touches zero, the saddle is a local minimum but has arbitrarily small eigenvalues, 
it is  {\it marginally stable}.
\begin{marginnote}[]
\entry{Marginal stable}{A stationary point with Hessian density of eigenvalues
vanishing continuously at zero.}
\end{marginnote}
\end{enumerate}
Below $T^* > T_d$ the free-energy landscape has a flat {\it threshold}~\cite{KuPaVi92},
a continuum of mar\-gin\-ally-\-stable states, which acts as an attractor
for the relaxation of random initial conditions at $T<T_d$~\cite{CuKu93}, as we will explain in Sec.~\ref{sec:relaxation-dynamics}.
At the threshold there are arbitrarily soft excitation modes that make the system extremely sensitive to small perturbations.
Typical stationary points are stable minima below the threshold and unstable saddles above it~\cite{CaGiPa98}.

This kind of rugged landscape is central to the (mean-field) theory of fragile glasses. 
Refined tools for counting, and
classifying by their stability, the local stationary points of highly non-convex landscapes, 
which build upon the Kac-Rice formula~\cite{Kac43,Rice44}, 
have been developed and extensively used in recent years~\cite{Ros19,Ros23,Auffinger23}. 
\begin{marginnote}[]
\entry{Threshold}{A marginally stable  sector of the free-energy landscape, above/below which stationary 
states are dominantly unstable/stable.}
\end{marginnote}

%In particular, the effect of a constant 
%field $\vec h$ with selected properties has been exploited to analyze these landscapes~\cite{Ros19}.

The equilibrium properties can be expressed in terms of 
sums over the states, counted with their multiplicity~\cite{DeDoYo}. This formulation ends up with  
%\begin{equation}
${\mathcal Z} = \int df \, e^{-N (\beta f - \sigma(\beta, f))}$
%\label{eq:DeDoYo}
%\end{equation}
where $\sigma = \Sigma/N$ is the complexity per degree of freedom at fixed randomness. The integral runs over 
free-energy densities and in the large $N$ limit it is dominated by the $f$ that minimizes the 
expression in the exponential, which could be the minimum $f$ if $\sigma=0$ or the non-trivial 
solution to $\beta  =\partial\sigma/\partial f$ otherwise. The latter case arises in the temperature interval 
$[T_s,T_d]$. 
%{\it Typical states} are those that  dominate the partition function. 
A {\it pinning field}
 method that allows one to select and study metastable states (which are not the 
 equilibrium ones) with this formalism was developed in~\cite{Monasson95}.

The equilibrium properties of the spherical $p\geq 3$ model at $T<T_s$ 
are also derived with a  replica analysis with a  {\it one step
replica symmetry breaking} (1RSB) {\it Ansatz} in which the elements of the 
Parisi matrix take two values, organized in square blocks. They represent the overlap between configurations
in the same state, $q_1=q_{\rm EA}$, and the overlap between configurations
in different states, $q_0=0$~\cite{CrSo92}. The size of the blocks is determined
by a parameter which is extremized. 
Instead, the properties at the threshold are obtained by fixing this parameter requiring marginal stability of the 1RSB {\it Ansatz}. 
Very recently, the static features of the spherical $p$-body model were re-derived with the 
{\it cavity method}~\cite{Gradenigo20}. 

Another way to access the non-trivial metastable states in the landscape is to define an 
{\it effective potential} for the overlap $q= N^{-1} \sum_{i} x^{(1)}_ix^{(2)}_i$ where $x^{(1)}_i$ and 
$x^{(2)}_i$ are two coupled copies of the system (same quenched randomness different configurations).
The logarithm of the probability distribution of $q$ defines the so-called {\it Franz-Parisi potential}~\cite{FrPa95}, 
\begin{marginnote}[]
\entry{Franz-Parisi potential}{
The free-energy of an equilibrium system constrained to have a fixed overlap with a   reference equilibrium 
configuration.}
\end{marginnote}
which at high temperatures has a single minimum at $q = 0$, 
indicating that the stable phase  is completely disordered, while as temperature is lowered,  loses convexity, and 
eventually develops a second minimum at $q \neq 0$ below $T_d$, indicating the presence and thermodynamic importance of 
metastable states.
 
 In the variant with bi-valued variables $x_i = \pm 1$, there is another transition at $T_G<T_s$, where $_G$ stands for Gardner~\cite{Gardner85}. 
The free-energy landscape takes on a hierarchical structure
and FRSB arises in the replica analysis below $T_G$. 
This transition separates a stable glass phase from a marginally stable one.
A sketch is shown in Fig.~\ref{fig:free-energy-sketches}(c).
This phenomenon also appears in particle based glassy models in infinite dimensions~\cite{Charbonneau14} and in some
of the ecological models~\cite{Altieri20} that we will discuss in Sec.~\ref{subsec:ecology}.

 \subsection{The mixed $p$ case}
 
We called pure $p$ model the one with a single monomial potential energy, Equation~\ref{eq:p-spin}. A logical extension 
is to consider the sum of two such terms with different $p$, say $p$ and $s$. There are several reasons for being interested in these 
generalizations. In the glassy context, they appear in the mode-coupling approach 
to super-cooled liquids. In mappings between optimization problems and disordered spin systems, 
models with ``polynomial'' energies naturally arise~\cite{Monasson97}.
These {\it mixed}  models turn out to
be much more complex than the pure ones, presumably because they lose the homogeneity of the Hamiltonian.
\begin{marginnote}[]
\entry{Temperature chaos}{
The TAP free-energy stationary points cross, merge or divide as temperature is 
varied.}
\end{marginnote}

Mixed models exhibit {\it temperature chaos}~\cite{Rizzo06}, that is, the 
finite temperature metastable states are not simple continuations of the zero temperature ones. Visualized as free-energy 
levels, there are level crossings~\cite{BaFrPa95,CaCaGiRi06,Crisanti06,Barbier20}. Moreover, 
the free-energy landscapes at fixed temperature are not layered as the ones of the monomial models. As the  
 free-energy landscapes are so involved, let us only describe them at zero temperature. First of all, 
they are different for the $(p=2)+s$ and $(p=3)+s$ models, and also 
diversify when $s$ is modified. In a few words, for the perturbed $p=2$ model the zero-$T$ replica solution 
 goes from RS ($s=0$) to 1RSB ($s=3$ and not too strong perturbation), followed by  a combined 
1RSB and FRSB structure, with the former/latter for large/small values of the overlap ($s>3$). At sufficiently strong perturbation it is just FRSB.
A similar (though different) complex phase diagram arises in the perturbed $p=3$ model~\cite{Auffinger22}.
Above the ground states, these models present an exponential number of local minima.
In the $p=3, s=4$ model there is an exponential number of marginally stable states 
in a {\it finite range of energies}~\cite{Folena20,Folena21} and not just a single value 
as in the pure model. All these novel features complicate considerably the 
dynamics, which is not completely understood yet, see Sec.~\ref{subsubsec:mixed}.

\section{RELAXATIONAL DYNAMICS}
\label{sec:relaxation-dynamics}

In this Section we focus on conservative forces $f_i=-\partial V/\partial x_i$, no external drive, and finite coupling to the bath 
($\eta\neq 0$).
%The coupling to the bath induces thermal fluctuations but also dissipation, and the consequent relaxation 
%of out of equilibrium initial conditions. 
Typical initial conditions of equilibrium at high temperature $T_0\gg T_d$ are very energetic and 
progressively transfer their excess energy to the bath when quenched to $T<T_0$. This process can be rapid and let the system 
equilibrate quickly with the environment, or very slow with peculiar time-dependencies.
Initial conditions which know about the free-energy landscape behave differently as we specify below.

In applications of these methods to the glass problem  $\gamma/m\gg 1$ 
and inertia is negligible. The velocities rapidly equilibrate with the 
environment and reach a Maxwell distribution. One then adopts an over-damped description from the 
start.
% and focuses on the behaviour of the coordinates $x_i$. 
%In quantum extensions the inertial term  must be present.

 \subsection{Slow relaxation and aging}

We focus on instantaneous temperature quenches to temperature $T$. Other protocols, like thermal annealing, 
 may be closer to the schemes used in industry and the like but we do not find it necessary 
to discuss them here.
In correspondence with the  free-energy landscape properties, the long-time relaxation 
of the $N\to\infty$ spherical pure $p$ models and their mixed extensions
have very different properties depending on $T_0$ and how $T$ compares to  various characteristic temperatures. 

\subsubsection{The  spherical pure $p=2$ case}

Quenches at the transition temperature $T=T_s$ show
critical dynamics. In sub-critical quenches, $T<T_s$,  from the high temperature phase, $T_0>T_s$, 
and no applied ordering field, $\vec h=\vec 0$,
%, the model yields a mean-field description of standard domain growth
the initial conditions tend to align in the 
course of evolution with $\vec v_N$ (or its reversed): 
$\langle \vec x \cdot \vec v_{N} \rangle \sim t^{3/4}$.
This parallels the growth of the zero wave-vector mode, homogeneous order,  
of finite dimensional  {\it coarsening}~\cite{CuDe95a}. 

\begin{marginnote}[]
\entry{Coarsening or domain growth}{
The progressive growth of order in the symmetry-broken phase across a second order phase transition.}
\end{marginnote}

The non-equilibrium character below $T_s$ is due to the fact that the system does not reach the required $N^{1/2}$ 
overlap with ${\vec v}_N$ in finite time with respect to $N$. This means that it does not 
achieve spontaneous symmetry breaking in such time scales. Having said so, the energy density 
gets close to its equilibrium value, 
%with the deviation decaying algebraically,  
$e(t) \sim e_{\rm eq} + c t^{-\alpha}$,
with  $e_{\rm eq}$  a function of $T/J$, 
$c$ a numerical constant and $\alpha=1$. 
In a coarsening problem, this decay estimates the growing  ``length''
%\begin{equation}
%$R(t) \sim  (\frac{e_{\rm eq}}{e(t) - e_{\rm eq}} )^{d-1} $
%\end{equation}
since the excess energy is localized on regular interfaces between domains.

The two-time correlation evolves in two two-time scales. A rapid and stationary one
\begin{eqnarray}
\lim_{t_2\to\infty} C(t_1,t_2) = C_{\rm st}(t_1-t_2)
\quad\mbox{with} \quad C_{\rm st}(0) = 1-q \;\;\mbox{and}   \lim_{t_1-t_2\to\infty}  \!\! C_{\rm st} (t_1-t_2) = 0
\; , 
\label{eq:limit-fast}
\end{eqnarray}
and a slow and {\it aging} one
\begin{marginnote}[]
\entry{Aging}{Older systems relax more slowly than younger ones. }
\end{marginnote}
\begin{eqnarray}
\lim_{t_1,t_2\to\infty} \!\! C(t_1,t_2) = C_{\rm ag}\left(\frac{t_1}{t_2}\right)
\;\; \mbox{with} \lim_{t_1\to t^+_2 \to\infty}  \!\! C_{\rm ag}\left(\frac{t_1}{t_2}\right) = q
\;\;\mbox{and} \lim_{t_1\gg t_2}  \!\! C_{\rm ag}\left(\frac{t_1}{t_2}\right) = 0
\; .
\label{eq:limit-slow}
\end{eqnarray}
The high-temperature initial conditions are completely forgotten asymptotically, in the sense that $C(t,0)\to 0$.
A sketch of this decay appears in Fig.~\ref{fig:sketch-corr-resp}(a), with the different curves corresponding 
to different $t_2$, increasing from left to right.

The time $t_2$ is the {\it age} of the system, the time spent at the working conditions before 
starting the measurement. In the long $t_2$ limit, 
a sharp separation in the decay of $C$ is achieved at $C=q>0$. This is the definition of the order parameter
given by Edwards and Anderson in their seminal paper on spin-glasses~\cite{MePaVi}. In this model, it coincides with 
the global equilibrium order parameter $q = q_{\rm EA} = N^{-1} \sum_{i=1}^N [\langle x_i\rangle^2] $ at the working temperature.
One can then propose the notion of {\it correlation scale}, there being 
two of them in this problem, $q\leq C<1$ and $0 \leq C<q$.
\begin{marginnote}
\entry{Correlation scale}{Range of correlation values with distinct properties from the rest.}
\end{marginnote}

The complete functional form $C_{\rm ag}(t_1/t_2)$ is known exactly. In analogy with coarsening, 
it is interpreted as evidence for an algebraic growth~\cite{Bray94} of the typical linear length of the domains $R(t) \sim t^{1/z}$, 
with a dynamic exponent $z$ which cannot be fixed in this case, since a change in $z$ would simply change 
the scaling function. Still, one can claim 
$C_{\rm ag}(R(t_1)/R(t_2))$ where times appear only through $R$, the
proxy for the typical length evaluated  at the two times. The essence of dynamic 
scaling is that all $n$-time correlations should be functions of $R(t_1)/R(t_2), \dots, R(t_{n-1})/R(t_n)$, a property which also holds in this model.

A similar separation applies to the integrated linear response function:
\begin{equation}
\chi(t_1,t_2) \equiv \int _{t_2}^{t_1} dt' \, R(t_1, t') 
\;\;\;\; \mbox{with} \;\;\;\;
\chi(t_1,t_2) \to \chi_{\rm st}(t_1-t_2)+ \chi_{\rm ag}(t_1,t_2)
\end{equation}
together with   $\chi_{\rm st}(0) = 0$,
\begin{equation}
 \lim_{t_1-t_2\to\infty} \chi_{\rm st}(t_1-t_2) = \frac{1}{T}  (1-q)
 \;  \;\;\;\; 
\mbox{and}
\;\;\;\;  
 \chi_{\rm ag}(t_1,t_2) = t_2^{-1/2} f_R\left(\frac{t_1}{t_2}\right) \xrightarrow[t_2\to\infty]{} 0 
 \; . 
\label{eq:FDT-st}
\end{equation}
In the rapid regime the correlation and linear response satisfy FDT
with the temperature of the bath. Instead, FDT is modified in the slow scale in a maximal way, 
as the integrated linear response vanishes asymptotically. 

In sub-critical quenches the separation of time-scales is 
{\it additive} as stationary and aging parts add up to yield the total two-time functions. A quench to the 
critical temperature would yield instead a {\it multiplicative} separation, 
$C=C_{\rm st} C_{\rm ag}$ and $\chi=\chi_{\rm st} \chi_{\rm ag}$, 
concomitant with the fact that $q=0$ 
at criticality. The relations between linear response and correlation functions in critical quenches have 
been studied in great detail~\cite{Calabrese-Gambassi}.
% and a comprehensive summary can be found in~\cite{Calabrese-Gambassi}.

An ordering field $h>0$ changes this picture: it introduces a finite time-scale $t_{\rm eq} = [(2+h^2)/\sqrt{1+h^2}-2]^{-1} \sim h^{-4}$ 
for $h\sim 0$, after which the dynamics become stationary and equilibrium is reached
even at low temperatures~\cite{CuDe95b}.

The properties exposed above are shared in full qualitative, and sometimes even quantitative, form 
by finite dimensional models of domain growth and phase separation, such as the two or three
dimensional Ising model with different kinds of stochastic microscopic updates. Examples 
and further details can be found in Refs.~\cite{Onuki02,Puri09,HenkelPleimling10}.

\subsubsection{The  spherical pure $p\geq 3$ case}

%\begin{enumerate}
%\item[(i)]
%\item

In this model there are three characteristic temperatures: 
the static critical 
temperature $T_s$, the dynamic critical temperature $T_d$, and $T^*$ the maximal temperature
at which metastable minima of the free-energy density exist. 

\paragraph{High temperature instantaneous quenches}

At $T>T_d$, initial conditions prepared in equilibrium at $T_0>T_d$
relax to the single  equilibrium state. 
After a short transient, $t_2$ \raisebox{-0.09cm}{$\stackrel{>}{\sim}$} $t_{\rm st}$, the dynamics 
become stationary,
%and both $C$ and $R$ depend on the time difference only, 
$C(t_1-t_2)$ and $R(t_1-t_2)$. At very high temperatures $T \gg T_d$ these functions 
quickly decay exponentially. Close to but still above $T_d$ the 
dynamics dramatically slow down. For all $T>T_d$, the energy density approaches the equilibrium value $e(t) \to e_{\rm eq}$
asymptotically. $T_d$ is the {\em dynamical} critical temperature below 
which all this no longer happens. 

\begin{figure}
%\hspace{1cm}
%(a) \hspace{3.5cm} (b) \hspace{3.75cm} (c)
%\\
\centerline{
\includegraphics[scale=0.4]{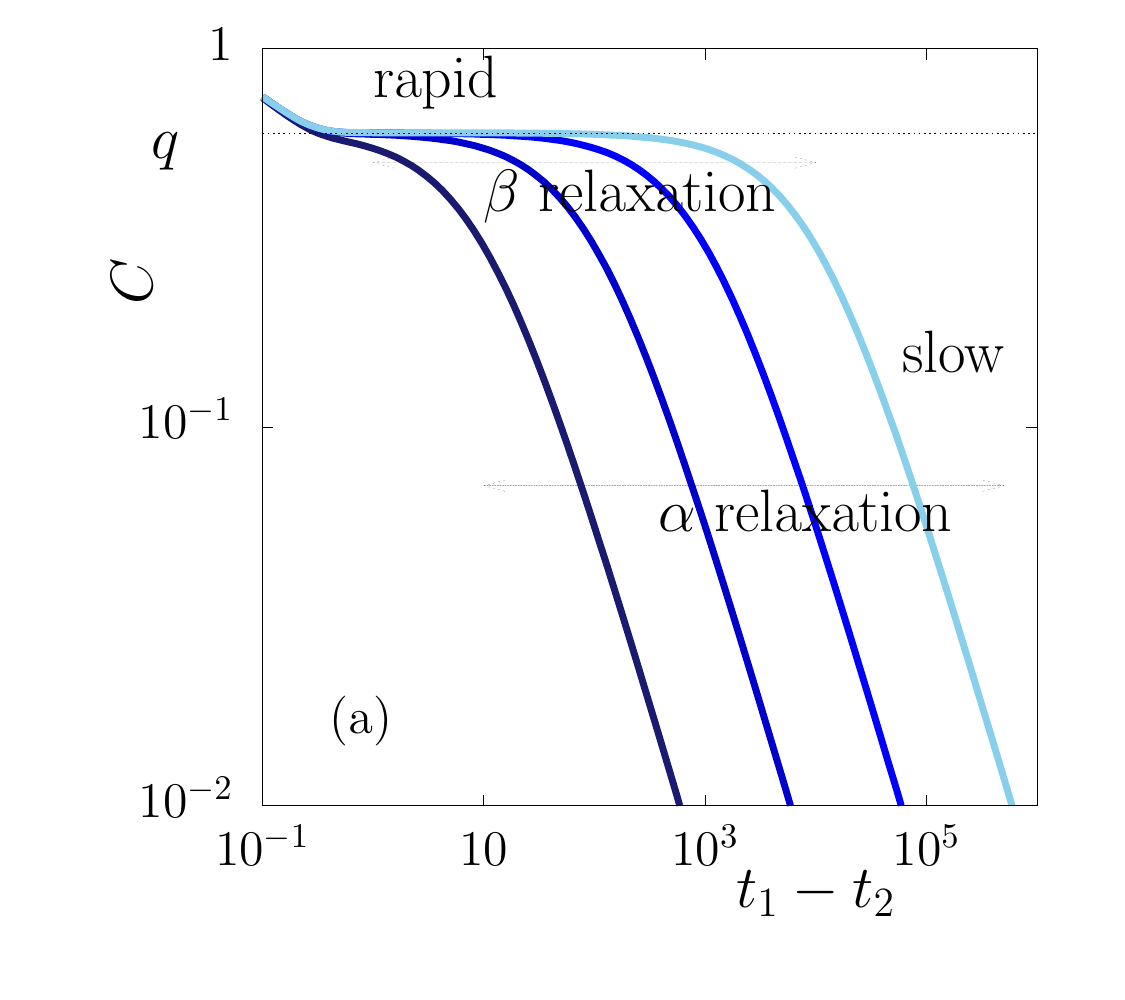}
\hspace{-9cm}
\includegraphics[scale=0.4]{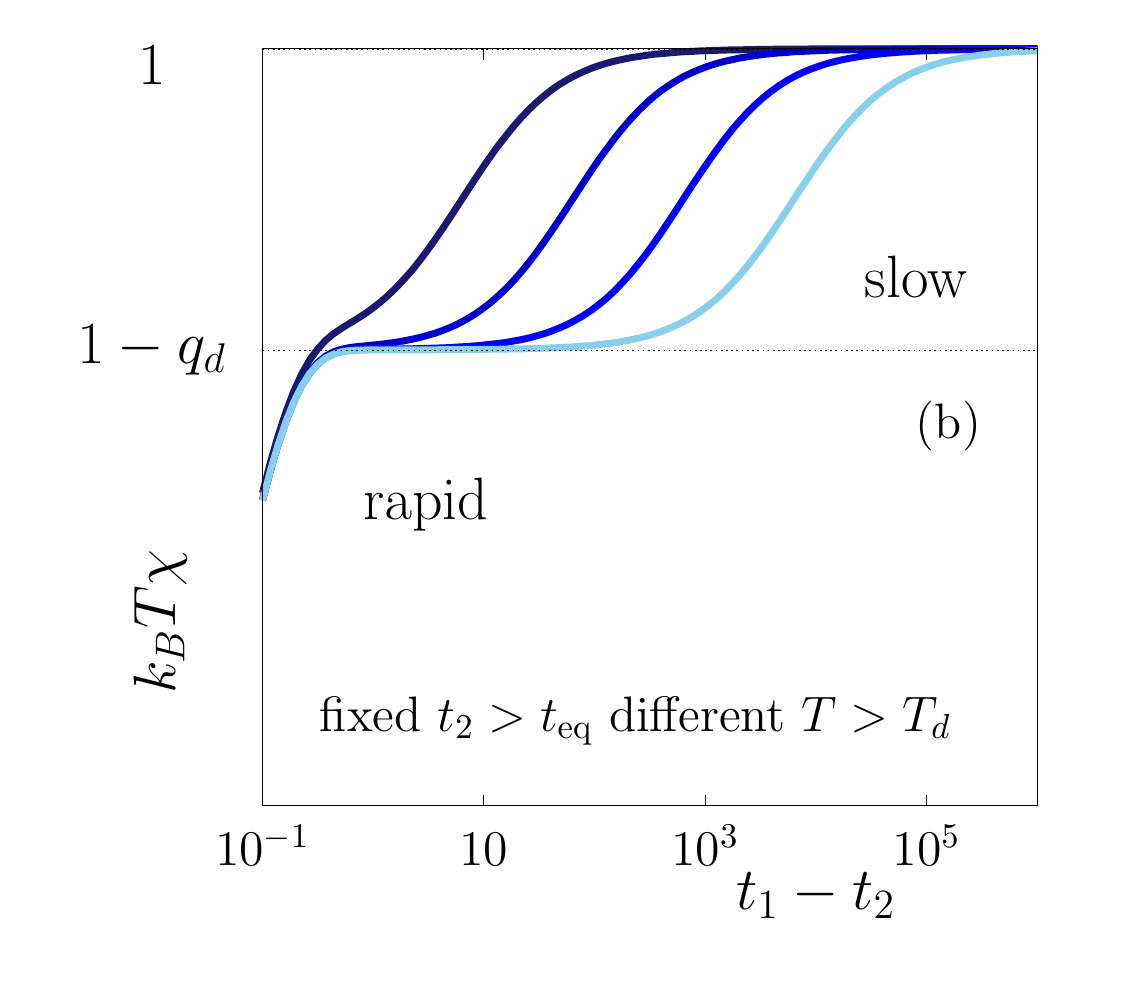}
\hspace{-9cm}
\includegraphics[scale=0.4]{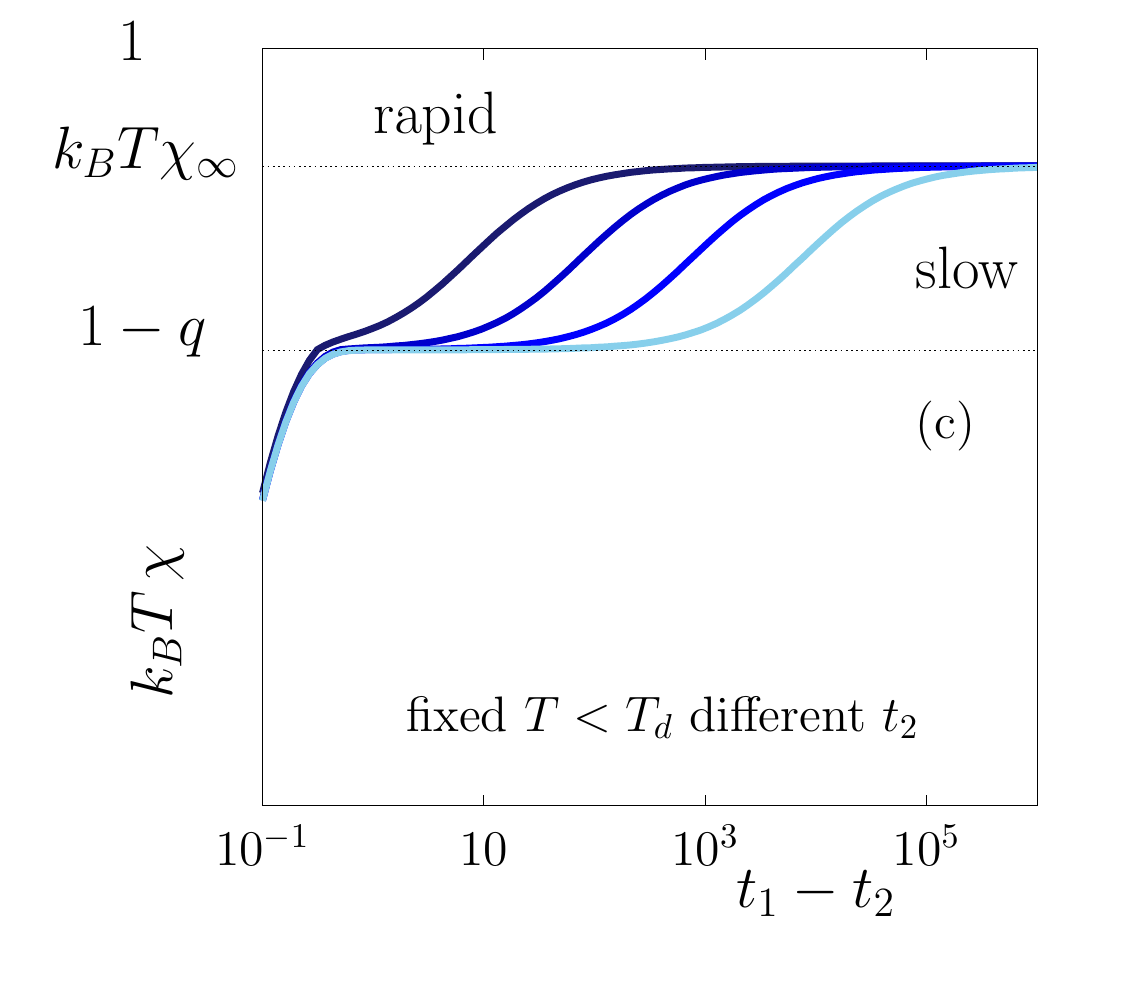}
}
\vspace{0.01cm}
\caption{(a) The two-time correlation $C(t_1,t_2)$. 
The sketch can represent different cases. First, the equilibrium decay 
at different temperatures $T$ getting closer to $T_d$ from above and from left to right
in the $p\geq 3$ model. The two-step relaxation is separated by a plateau at $q=q_d$. 
Second, the out of equilibrium decay below $T_d$ in the $p\geq 3$ model, or below $T_d=T_s$ in the $p=2$ one, with
the different curves calculated for increasing waiting times $t_2$ from left to right. 
 (b) The integrated linear response times working temperature, $k_BT \, \chi(t_1,t_2)$ in the equilibrium $p\geq 3$ model 
 at different temperatures  close to $T_d$, decreasing from left to right (we restored $k_B$).
 (c) The integrated linear response $\chi(t_1,t_2)$ out of equilibrium at a single temperature $T<T_d$ and 
 for different $t_2$ increasing from left to right.}
 \label{fig:sketch-corr-resp}
\end{figure}

Close to $T_d$  a two-step relaxation develops: 
a first rapid decay from $C(0)=1$ towards a plateau at  $q_d$, 
a slow evolution around it and, eventually, a further very slow decay towards zero. 
The plateau length is longer and longer as $T$ gets closer to $T_d$, representing the 
fact that the system keeps an increasingly long memory of the initial configuration, which it will 
eventually forget.
These features are typical of {\it super-cooled liquids}.
\begin{marginnote}[]
\entry{Super-cooled liquids}{
A liquid cooled beyond the crystallization point.}
\end{marginnote}
The evolution around $q_d$  is  the {\it $\beta$-relaxation} and it is 
algebraic with two exponents, 
$C(t_1-t_2) \sim q_d + c (t_1-t_2)^{-a}$ and $C(t_1-t_2) \sim q_d - \overline{c} (t_1-t_2)^{b}$, 
related by $\Gamma^2(1+b)/\Gamma(1+2b) = \Gamma^2(1-a)/\Gamma(1-2a) =
(T/2) \,  \overline V'''(q)/ (\overline V''(q))^{3/2}$. 
The {\it structural} or {\it $\alpha$-relaxation} below the plateau, $C(t_1-t_2) = f((t_1-t_2)/\tau_\alpha)$,
can be characterized in terms of a {\it structural relaxation time}
which diverges as $\tau_\alpha \sim |T-T_d|^{-\gamma}$ with $\gamma$ a ``critical'' exponent satisfying 
$\gamma=1/(2a) + 1/(2b)$. At infinite time separation the decorrelation is complete, $f(\infty)=0$.
The sketch in Fig.~\ref{fig:sketch-corr-resp}(a) represents this case, with  the curves at
different temperatures decreasing from left to right.

In terms of trajectories on the spherical configurational space, the interpretation is the following. Let us assume that $t_1\geq t_2 \gg t_0$.
At $t_1$ such that $t_1-t_2$ is short, the position vector $\vec x(t_1)$ remains close to the reference one at 
$t_2$, $\vec x(t_2)$, and $C(t_1-t_2)>q_d$. Instead, at longer  $t_1-t_2$, 
$\vec x(t_1)$ drifts towards other equilibrium configurations, $C(t_1-t_2)<q_d$, and eventually decorrelates completely from 
the reference one, $C(t_1-t_2) \to 0$.

The integrated linear response $\chi$ also develops an additive  separation of scales, 
depicted in Fig.~\ref{fig:sketch-corr-resp}(b). The correlation plateau at $q_d$ translates
into a plateau at $(1-q_d)/T$ in $\chi$.
Correlation and linear
response are related by the FDT: $T \, R(t_1-t_2) = -d_{t_1} C(t_1-t_2)$, for $t_1-t_2\geq 0$,
for all $t_1-t_2$.  

\paragraph{Low temperature instantaneous quenches}

Below $T_d$, initial conditions prepared in equilibrium at $T_0>T_d$ 
reduce their  energy until reaching the threshold energy which is macroscopically higher than the equilibrium 
one, $e(t) \to e_{\rm th} + c t^{-\alpha}$
with $e_{\rm th} > e_{\rm eq}$. 
The relaxation takes place in two two-time scales, say, 
$(t_1-t_2)/\tau_\alpha(t_2) < 1$ and $(t_1-t_2)/\tau_\alpha(t_2) > 1$. In the first one, $C$ and $R$ 
behave as if the system were equilibrated with the environment, 
they are stationary and satisfy FDT. In the second one, 
they are non-stationary with {\it aging}, that is, older systems decay more slowly than younger ones.
This represents the fact that the system keeps exploring  phase space with a velocity that 
diminishes with the elapsed time after the quench. The physical aging of glasses has been studied
experimentally during the last century since the implications on the material properties have 
industrial relevance. 
A recent non-technical discussion can be found in Ref.~\cite{Middleton22}.
In the slow regime, the FDT applies but with a different temperature from the one of the bath~\cite{CuKu93}; this 
is related to the fact that the threshold energy level is asymptotically sampled uniformly, hence attaining a sort of 
{\it effective equilibrium} at an {\it effective temperature}~\cite{CuKuPe97}. 
For $T_0>T_d$, $T_{\rm eff}>T$, recalling the fact that the system has been quenched from high temperature and 
the long time-delay dependencies (low frequencies) have not had enough time to equilibrate with the 
bath. The operationally defined $T_{\rm eff}$ then 
meets the phenomenological idea of {\it fictive temperatures} commonly used in the glass literature~\cite{Scherer86}.
\begin{marginnote}[]
\entry{Effective temperatures}{
The slow dynamics self-organizes to take place at a temperature that is different from the one of the
bath, and has a thermodynamic meaning.}
\end{marginnote}
\begin{marginnote}[]
\entry{Weak ergodicity breaking}{
The approach to the plateau suggests ergodicity breaking; however, there is no 
full arrest and $C\to 0$ asymptotically.}
\end{marginnote}

The separation between scales is sharp in the infinite {\it waiting-time} limit:
\begin{eqnarray}
&& 
\lim_{t_2\to\infty} C(t_1,t_2) = C_{\rm st}(t_1-t_2)
\qquad\qquad\quad
\mbox{with} \qquad \lim_{t_1-t_2\to\infty} C_{\rm st}(t_1,t_2) = 0
\; , 
\label{eq:limit-fast}
\\
&&
\lim_{t_1,t_2\to\infty} C(t_1,t_2) = C_{\rm ag}({\rm h}(t_1)/{\rm h}(t_2))
\qquad\mbox{with} \qquad \lim_{t_1\to t^+_2 \to\infty} C_{\rm ag}(t_1,t_2) = q
\; ,
\label{eq:limit-slow}
\end{eqnarray}
together with $C_{\rm st}(0)=1-q$ and  $\lim_{t_1\gg t_2} C_{\rm ag}({\rm h}(t_1)/{\rm h}(t_2)) = 0$.
These properties have been named {\it weak ergodicity breaking}~\cite{Bouchaud92,CuKu93}.
The numerical solution of the Schwinger-Dyson equations suggests that ${\rm h}(t) \propto t$~\cite{Kim,Folena20}.
The $\beta$ relaxation around  the plateau is also controlled by two exponents, 
$C(t_1-t_2) \sim q_d + c (t_1-t_2)^{-a}$ and $C(t_1,t_2) \sim q_d - \overline{c} ({\rm h}(t_1)/{\rm h}(t_2))^{b}$, 
related by $(T/T_{\rm eff}) \, \Gamma^2(1+b)/\Gamma(1+2b) = 
\Gamma^2(1-a)/\Gamma(1-2a) = (T/2) \,  \overline V'''(q)/ (\overline V''(q))^{3/2}$~\cite{CuLe96}. 

\begin{figure}
%\hspace{1cm}
%$(a)$ \hspace{3.5cm} $(b)$ \hspace{3.75cm} $(c)$ 
%\\
\centerline{
\includegraphics[scale=0.38]{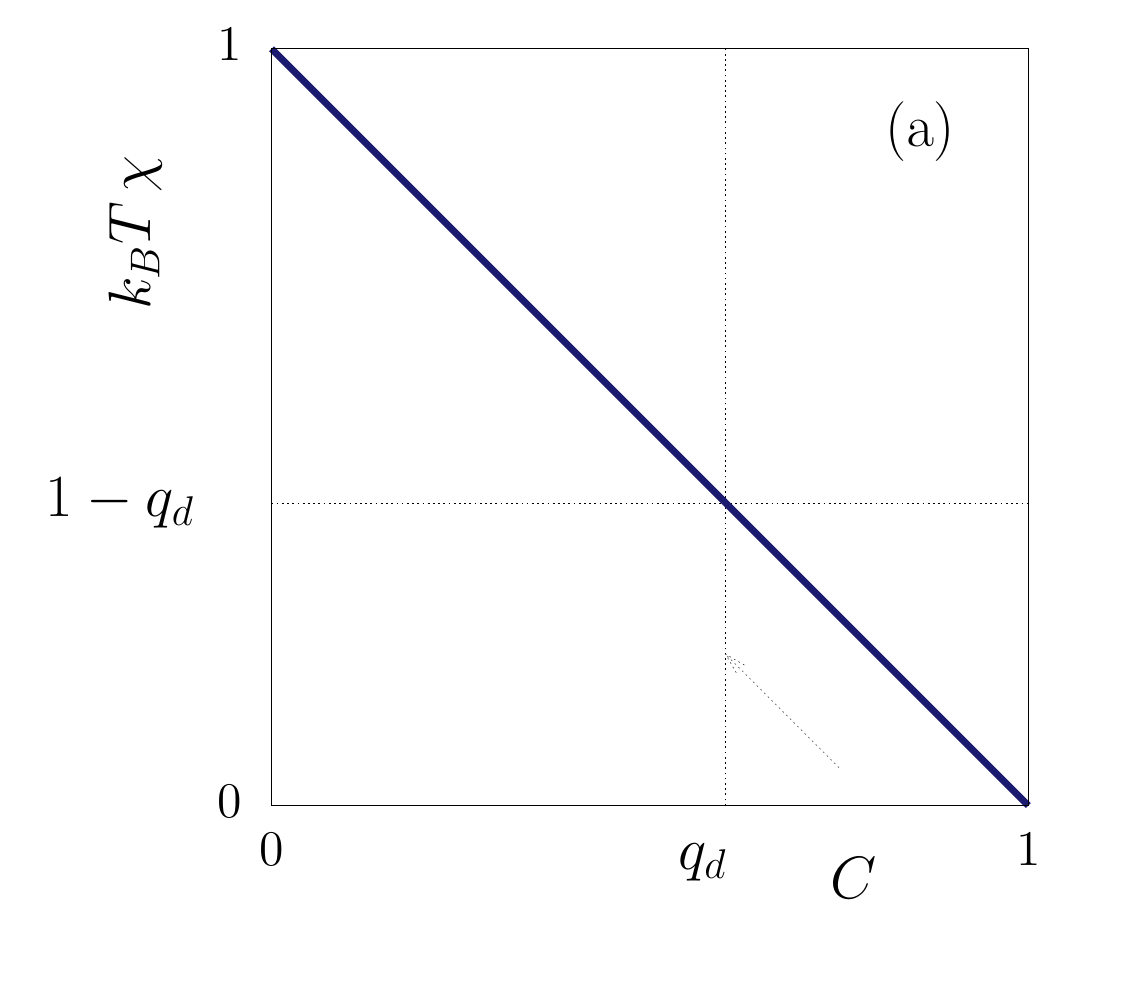}
\hspace{-9.1cm}
\includegraphics[scale=0.38]{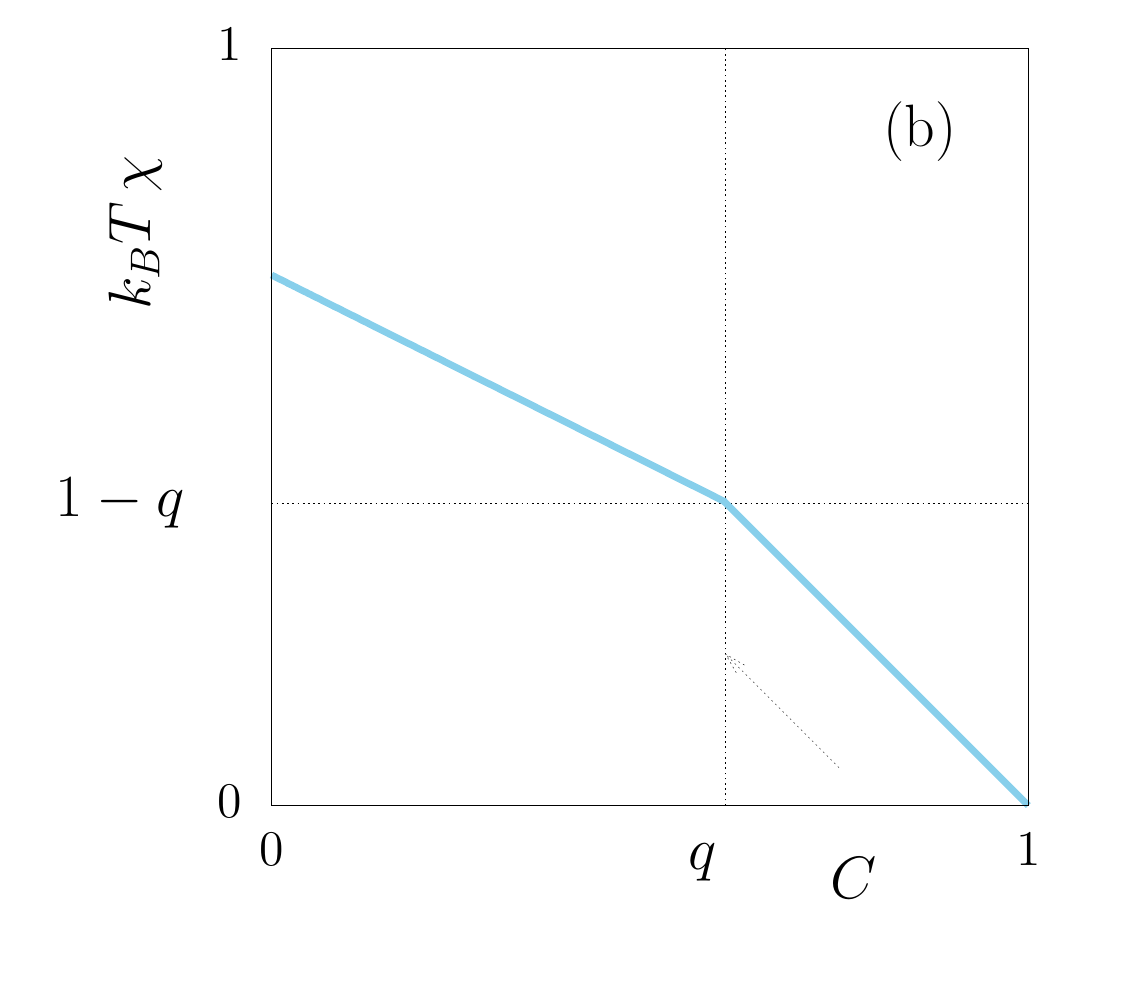}
\hspace{-9.1cm}
\includegraphics[scale=0.38]{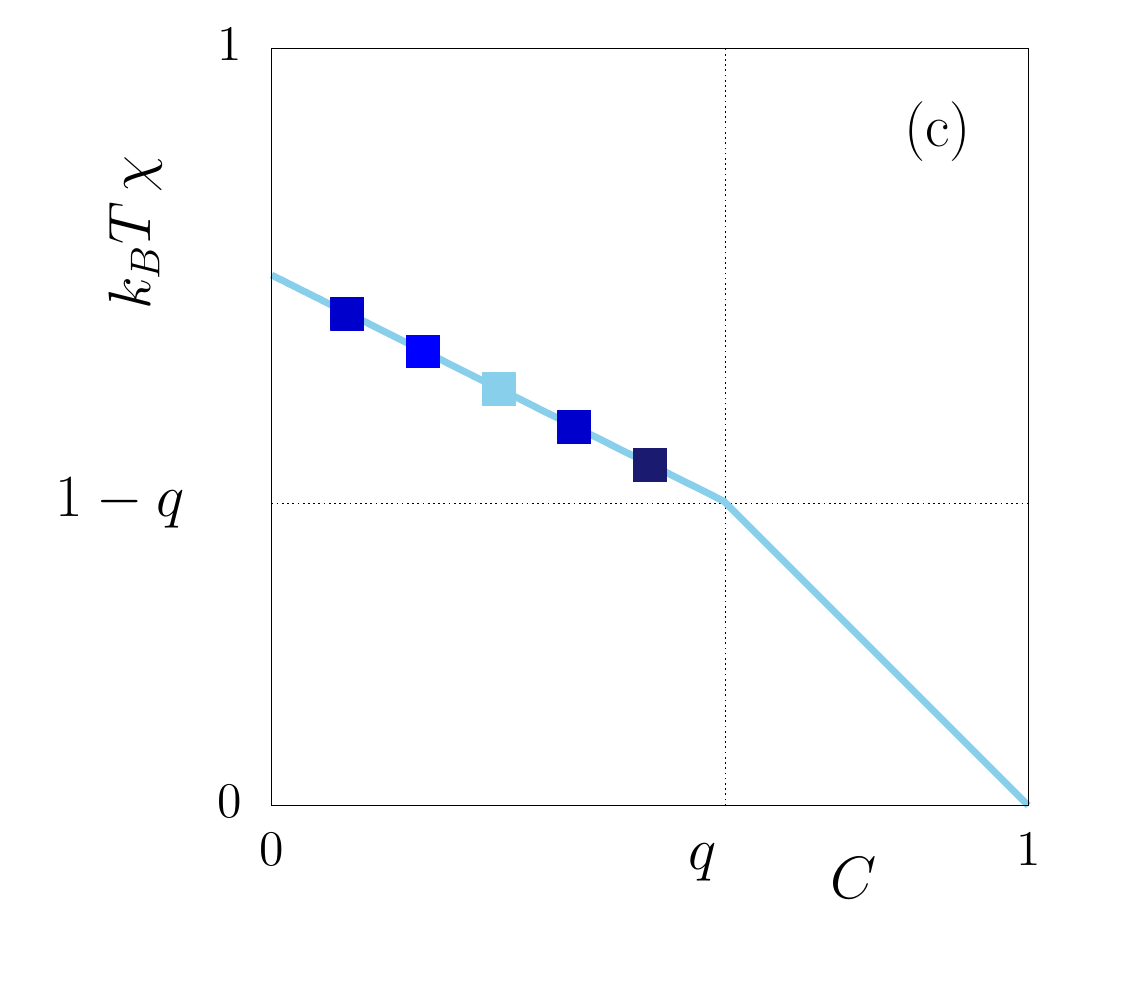}
}
\vspace{0.01cm}
\caption{The parametric construction $\chi(C)$. Both axes are parametrized by $t_1-t_2$ which runs from 
0 ($C=1, \chi=0$) to $\infty$ ($C=0, \chi=\chi_\infty$). (a) In any equilibrium situation. 
If close but above $T_d$ in models with a RFOT, $q_d$ indicates the value of $C$ on the plateau. (b) 
Out of equilibrium below $T_d$ in the $p\geq 3$ model. The dotted arrows in (a) and (b) 
indicate the sense in which $t_1-t_2$ increases. 
(c) Thermal noise induced fluctuations on case (b). The data points are obtained at fixed $t_1$ and $t_2$. The 
skyblue point represents the values averaged over thermal noise while the four dark blue points
would be the results of retarded or advanced runs with respect to the average, constrained to lie on the 
$\chi(C)$ curve following time-reparametrization  $t\mapsto {\rm h}(t)$.}
\label{fig:parametric}
\end{figure}

While the linear response vanishes at long-time delays, 
\begin{equation}
\lim_{t_1\gg t_2} R(t_1,t_2)=0
\; , 
\end{equation} 
the  integrated linear response 
$
\chi(t_1,t_2) = \int_{t_2}^{t_1} dt \, R(t_1,t)
$
does not, an effect called {\it weak long-term memory}~\cite{CuKu93}. 
A sketch with the integrated linear response for increasing $t_2$ from left to right 
is displayed in Fig.~\ref{fig:sketch-corr-resp}(c).
Moreover,
\begin{eqnarray}
&& 
 \lim_{\substack{t_2\to\infty\\
C(t_1,t_2)=C}} \chi(t_1,t_2) = \chi(C)
= 
\left\{
\begin{array}{ll}
T^{-1} \, (1-C) \qquad & q \leq C\leq 1
\\
T^{-1} \, (1-q)
%{\rm ct} 
+ T_{\rm eff}^{-1} \, (q-C) \qquad & 0 \leq C< q
\end{array}
\right.
\label{eq:qFDT}
\end{eqnarray}
%with ct = $T^{-1} \, (1-q)$, 
similarly to Equation~\ref{eq:FDT-st},
and $T_{\rm eff}>T$.
\begin{marginnote}[]
\entry{Weak long-term memory}{The linear response integrated over a 
diverging period is finite but the instantaneous one vanishes at long time delays. }
\end{marginnote}

This piece-wise relation is a particular case of the more general parametric one
\begin{equation}
 \lim_{\substack{t_2\to\infty\\
C(t_1,t_2)=C}} \chi(t_1,t_2) = \chi(C) \qquad\mbox{with} \qquad \frac{1}{ T_{\rm eff}(C)} = \chi'(C)
\; .
\end{equation}
%\item[(iii)]

In terms of evolution in  the free-energy landscape, the highly energetic initial configurations approach the 
threshold  level and keep straying along it without penetrating below it in finite times with respect to system size. 
 The threshold is the basin of attraction of {\it all} high temperature 
initial conditions~\cite{Folena20}. The  decay of the correlation in two two-time scales 
indicates that the threshold is made of almost flat ``channels'' which   trap the system in ``transverse'' directions
but let it drift in ``longitudinal'' ones. 
Thus, the correlation first decays from $C(t_1,t_1)$  to a maximal decorrelation $q=q_{\rm th}$, the transverse 
size of the channels, for 
$t_1-t_2\to\infty$. The value $q_{\rm th}$ coincides with $q_{\rm th} = N^{-1} \sum_i [\langle x_i\rangle^2]$
where $\langle x_i\rangle$ are the local order parameters of the threshold free-energy level.
Then, $C$ further decreases from $q_{\rm th} $ to zero in an aging manner, since getting closer and closer to completely flat 
directions~\cite{CuKu93,Kurchan96}. 

\begin{figure}[b!]
\centerline{(a) \hspace{5cm} (b) \hspace{1.5cm} (c) \hspace{1.5cm} (d)  }
\centerline{
\includegraphics[scale=0.2]{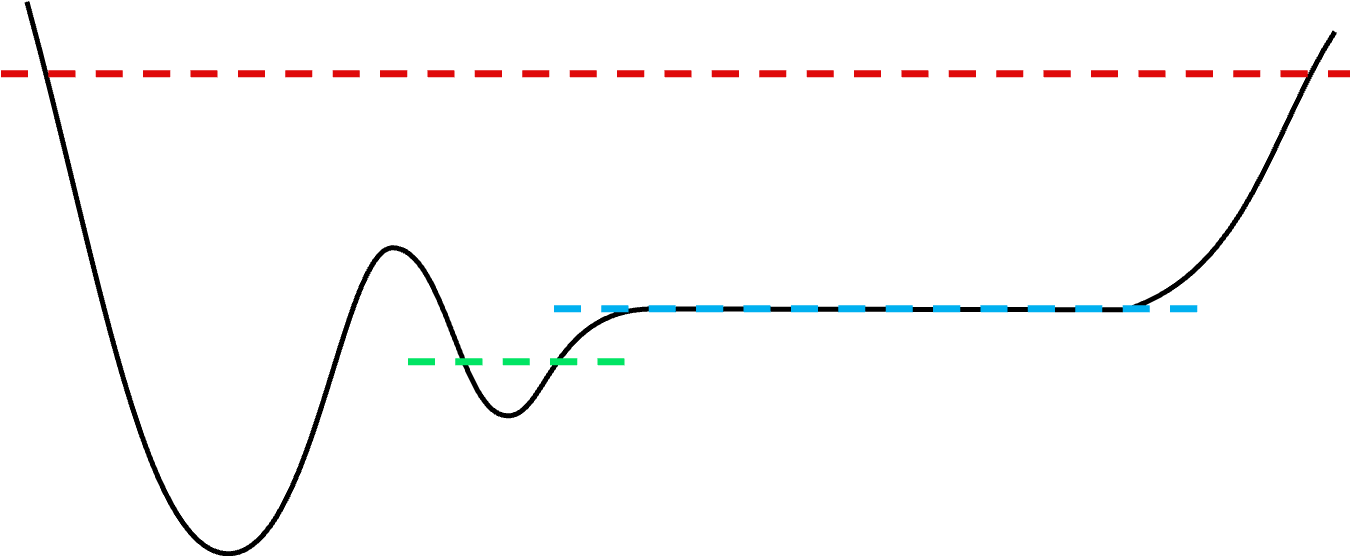}
\hspace{-7.5cm}
\includegraphics[scale=0.2]{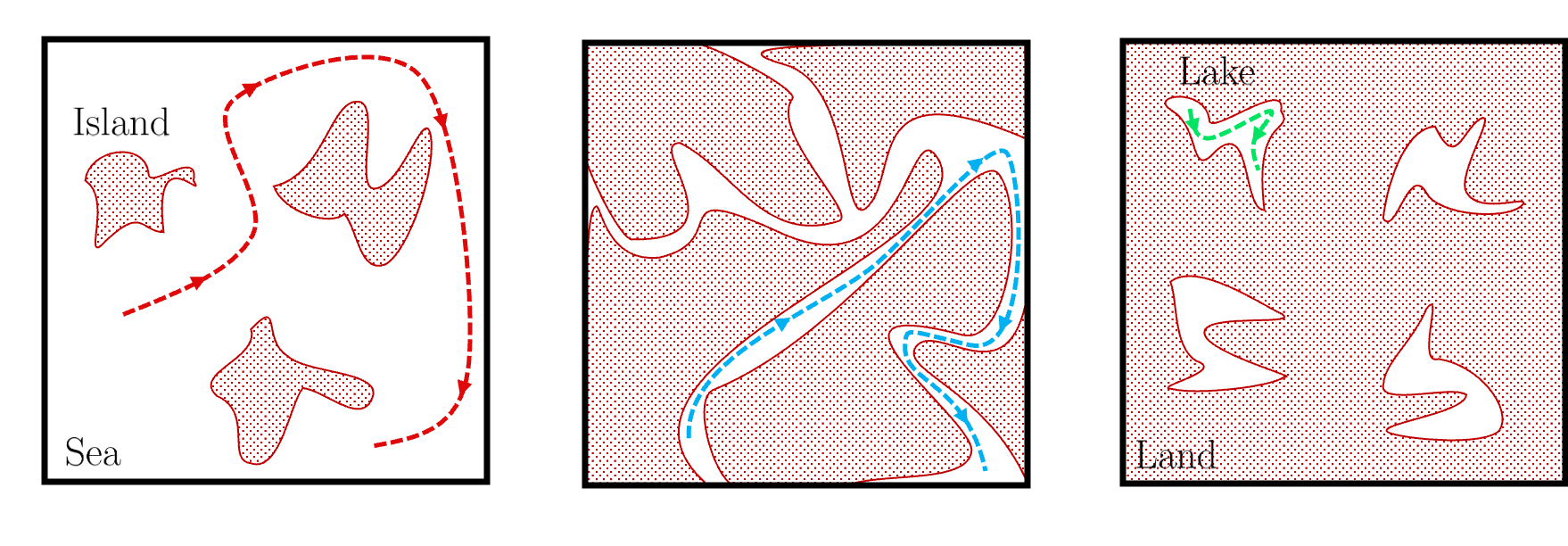}
}
\vspace{0.5cm}
\caption{A sketch of the dynamics trajectories in the free-energy density 
landscape of the $p\geq 3$ spherical model.  $(a)$ A ``transverse'' cut, with a level above threshold  (dashed red), 
at threshold (dashed blue and below threshold  (dashed green). The free-energy density is defined on an $N$ 
dimensional space and $N-1$ coordinates are therefore hidden in this sketch. The threshold is mostly 
flat while the sub-threshold states are mostly stable.
Three top views of the landscape taken at different free-energy levels: above $(b)$, at $(c)$ and below $(d)$ the threshold. 
The ``islands'' represent barriers protruding above the selected level. The ``sea'' represents the available space for the 
system to move or relax. The threshold is visualized as being very ramified and expanding along phase space.}
\label{fig:free-energy-landscape-p3}
\end{figure}

The dynamics close and below $T_d$ have the qualitative features 
of the (schematic) Mode Coupling Theory (MCT) description of the slowing down of fragile glasses~\cite{Bouchaud96,BeBi11,Ca09}.
There is a technical reason for this. 
The introduction of quenched randomness exactly renders all diagrams from the field-theoretic solution sub-leading 
except for the melonic ones~\cite{Bouchaud96}. The idea of ``simplifying'' the underlying model by introducing randomness, 
follows the same logical path taken by Kraichnan in the formulation of the direct interaction approximation 
for the Navier-Stokes equation~\cite{Kraichnan59}. 
\begin{marginnote}[]
\entry{Mode Coupling Theory}{Approximate self-consistent equations ruling the evolution of 
selected dynamical correlations in interacting particle models, {\it e.g.} 
the intermediate scattering function.} 
\end{marginnote}

\paragraph{Instantaneous quenches within metastable states}

The evolution, below their spinodal, of initial conditions prepared in one of the metastable states 
remains confined within this same state~\cite{FrPa95,BaBuMe96}.  
This confirms the fact that the barriers surrounding these states diverge with $N$ and cannot be surmounted in times that do not scale conveniently with $N$.
After a short transient the dynamics become stationary. The overlap of the time-dependent configuration with the initial one saturates to a non-zero value, 
$C(t_1-0) \to \overline q_1$. 
Accordingly, also $\lim_{t_1\gg t_2 \gg t_0} C(t_1-t_2) \to q_1>0$. These two overlaps characterize the similarity of the 
initial state with the final one, $\overline q_1$, and the size of the final state, $q_1$. 
The FDT holds, reflecting the fact that the system achieves a {\it restricted equilibration} within the chosen metastable state, 
in the manner of conventional {\it ergodicity breaking}. 
%However,  there is no full equilibration since all states cannot be sampled  if the long time 
%limit is taken after the thermodynamic limit.
\begin{marginnote}[]
\entry{Strong ergodicity breaking}{The decorrelation is not complete. }
\end{marginnote}

%\end{enumerate}

\paragraph{A pictorial view}

Figure~\ref{fig:free-energy-landscape-p3} depicts graphically the different situations discussed above 
in more abstract terms. The panels (b)-(d) represent a top view of the free-energy 
landscape and how it can be explored in quenches from $T_0>T_d$ to $T>T_d$~(b), 
$T_0>T_d$ to $T<T_d$~(c), and the dynamics within confining metastable states $T_0<T_d$~(d).

\subsubsection{The spherical mixed $p$ case}
\label{subsubsec:mixed}

The relaxation of the spherical {\it mixed} $p-s$ models is more complex and still the subject of active research.  

The analysis of the state following dynamics in the $p=3$, $s=4$ case, for $T_s<T$ $< T_0<T_d$, was carried out in~\cite{BaFrPa95,CaCaGiRi06,KrZd10a,KrZd10b,SuCrKrLeZd12}. 
Above a special working temperature the relaxation is as in the pure 
model, in the sense that the system equilibrates within the transformed original state and $C(t,0)$ approaches a 
non-zero value consistent with metastable (time-independent) calculations. 
%Surprisingly enough, 
Below it, 
these solutions no longer exist. Instead, the 
dynamics ages forever with the very unusual feature of keeping memory of the initial condition, via a non-zero
asymptotic $C(t, 0)$, and thus realizing {\it strong ergodicity breaking} combined with aging. Below this 
special temperature the initial state opens up into a marginal region of the free-energy landscape, 
which lies {\it below} the level approached with quenches from
$T_0\to\infty$. The configurations in these new marginal states do keep, however, a finite similarity with the initial ones. 

More recently, these  results were complemented 
by the numerical study of the gradient descent ($T=0$) dynamics of initial conditions thermalized 
close~\cite{Folena20} and far above~\cite{Folena23}  $T_d$, for a wide variety of $s$ terms added to both $p=2$ and $p=3$.
By numerical we mean the numerical solution of the Schwinger-Dyson equations. Differently from the monomial case, 
the asymptotic configurations seem to keep memory of the initial ones, $\lim_{t\to\infty} C(t,0) >0$, 
for all $T_0$ and, presumably, for all $s \neq  p$. The dynamics remain confined to a restricted
manifold that depends on the initial condition. Moreover,  the asymptotic energy is not  the threshold one at which dominant 
minima become saddles (as in the pure case), but a higher value, which depends on $T_0$. This 
level is also  a marginally stable region of the potential energy landscape and, consequently, the 
energy density (and other one-time quantities) converge algebraically to their asymptotic values.
Therefore, in these models there is no unique basin of attraction for 
 {\it all} high-temperature initial conditions. Concerning the scaling of  $C$ and $\chi$, 
 the data suggest that they are both functions of $t_1/t_2$.  $\chi(C)$ takes a non-trivial functional form, 
 thus invalidating the identification of simple aging with a single $T_{\rm eff}$ value.

The features explained above were obtained numerically and are  limited in precision and time
span, so changes in the truly asymptotic regime cannot be excluded. Unfortunately, no analytic 
solution, compatible with  the numerical results, has been derived yet for the mixed case.
While a connection between 
dynamic and metastable behaviour derived from replica calculations 
was clear in monomial models, this is still lacking for the  $p+s$ cases.

\subsubsection{Models in the Sherrington-Kirkpatrick class}

The Sherrington-Kirkpatrick model, as already stated, belongs to a different equilibrium class 
 (FRSB, hierarchical organization of equilibrium states with all possible overlaps between them) 
and this has a dynamic correspondence.  The out of equilibrium
Langevin relaxation of soft continuous spins~\cite{Sompolinsky81,SompolinskyZippelius81}, 
after a quench from $T>T_s$,
occurs in an infinite sequence of times-scales~\cite{CuKu94,FrMe94}. 
The FD relation is  linear $T\chi(C) = 1-C$
for $C>q$ and takes a non-trivial form for $C<q$. 

In its original strictly Ising version other algorithms can be developed to search for the configuration that corresponds to the 
optimum energy, 
see~\cite{Montanari20,Montanari21} for recent progress.

 \subsection{Effective temperatures and multi-thermalization}
 
 The relaxation of the monomial  model takes place in two scales, each 
 with its own temperature: the fast one is the one of the  external bath while the
 slow scale arranges to have its own value self-consistently selected 
 by the initial condition and the internal interactions.
 There are other models (e.g. the Sherrington-Kirkpatick or the $p$-spin in the 
 Gardner phase, both with  soft variables) for which the relaxation 
 follows a sequences of scales, each 
 with its own temperature~\cite{CuKu94,FrMe94,Altieri20}. 
All these cases are included in  
% \begin{equation}
$- 1/T_{\rm eff}(C)= \chi'(C) $, 
%\; . 
% \end{equation}
in the long $t_2$ limit.
 This definition has a thermodynamic meaning explained in~\cite{CuKuPe97,CuKu00} 
 and later extensively investigated~\cite{Cu11}.
%In the fully connected models that we focus on here it can be put in contact with the 
%replica symmetry breaking parameter, called $x$ or $m$ in that formalism
$T_{\rm eff}$  can be accessed with 
 thermometers which tune their measurement to the desired correlation scale.
% In the long term or weak entropy production limits, 
 The degrees of freedom evolving in the same scale are 
 equilibrated among them
 and share the same value of $T_{\rm eff}$. Instead, the ones which evolve in different 
 scales are not and can have different temperatures.
 This same scenario can be induced on, {\it e.g.}, a harmonic oscillator by 
 coupling the particle's position to a multi-bath with different temperatures and 
 time-scales~\cite{CuKu00,CoCoKuMi21} (different pairs $\Gamma_R$ and $\Gamma_C$ related by FDTs at different temperatures). 
 In the case in which the baths evolve in 
 widely separated time-scales their effects can be mimicked by quasi-static random fields~\cite{CuKu00}.
 This formalism has been recently exploited~\cite{Altieri20} to recover the 
 aging with effective temperature scenario from the single effective variable
 formalism. 
 %Interestingly enough, the effective temperature of the
%slow degrees of freedom is fixed by requiring critical dynamics
%on short time-delay scales, i.e. marginality.
The effective temperature also has an intuitive meaning. For quenches from disordered/extended 
initial conditions to low temperature phases $T_{\rm eff}(C<q) \geq T$.
For inverse quenches in which an ordered/localized initial configuration is evolved
under conditions that render it more disordered, $T_{\rm eff}(C<q) \leq T$. Accordingly, the slow 
scales, $C<q$, remember the initial conditions and their ``more ordered'' or ``more disordered'' 
nature, compared to the target one at the running temperature.
In $p$ body models $T_{\rm eff}$ also finds a fascinating relation with the structure
of the free-energy density explored by the dynamics:
\begin{equation}
\frac{1}{T_{\rm eff}(C<q)} = \left. \frac{\partial \sigma(f,T)}{\partial f}  \right|_{f_{\rm th}}
\!\!\!\! .
\end{equation}
with $\sigma$ the complexity per degree of freedom~\cite{Monasson95}. 
The effective temperature concept, the connection with phenomenological ideas, and its measurement
in a large variety of finite $d$ and systems has been reviewed in Ref.~\cite{Cu11}. More recent applications in the 
context of learning in neural networks will be discussed in Sec.~\ref{subsec:acceleration}.

 \subsection{Time reparametrization invariance}
 \label{subsec:time-rep-inv}
 
 A very general feature of the slow dynamics is that the time-dependence of $C$ and $R$ is sensitive
to vanishingly small changes in the equations of motion. Take for example the case of 
ferromagnetic coarsening. An arbitrary small random field changes the
growth of the equilibrated domains from algebraic to  logarithmic. 
In mean-field disordered systems  
weak non-conservative forces may destroy the aging relaxation and 
render the evolution stationary (Sec.~\ref{sec:driven-dynamics}). This extreme sensitivity is the consequence of the 
slow relaxation taking place along flat regions of  the free-energy landscape. These
facts have a mathematical counterpart in the invariances of the Schwinger-Dyson
equations.
 
 In the absence of an external force $\vec h=\vec 0$, the low temperature evolution of the $p\geq 3$ models 
 acquires a (global) time-reparametrization invariance~\cite{SompolinskyZippelius81,CuKu93,Sompolinsky81,Ioffe88,Chamon02}:
 \begin{equation}
 t \mapsto {\rm h}(t) 
\; , 
\;
\quad
\;
C(t_1,t_2) \mapsto C({\rm h}(t_1),{\rm h}(t_2))
\; , 
\;
\quad
\;
R(t_1,t_2) \mapsto \frac{d{\rm h}(t_2)}{dt_2} \, C({\rm h}(t_1),{\rm h}(t_2))
\; . 
\label{eq:t-rep}
 \end{equation}
Under these transformations, 
certain relations between observables and, in particular, the ones between 
linear response and correlation,  remain unmodified. For example, 
%\begin{equation}
$
T \chi(t_1,t_2) \; \equiv \; T \, \int_{t_2}^{t_1} dt \, R(t_1, t) = 1-C(t_1,t_2)$
for 
$
t_1-t_2 >0
$
%\end{equation}
is still 
%\begin{equation}
$
T  \chi(C) = 1-C
$
%\end{equation}
for any way of measuring time, be it as $t$ or any other monotonic ${\rm h}(t)$. This feature extends to any 
relation between $\chi$ and $C$ of the form $\chi(C)$.

The time-reparametrization invariance implies that, in the 
asymptotic long times limit, there exists a family of solutions to the Schwinger-Dyson  equations 
 linked to each other by one of the transformations in  Equation~\ref{eq:t-rep}. 
This is a consequence of the fact that the proper measure of ``time distance between two times 
$t_1$ and $t_2$'' is the value of the 
correlation $C(t_1,t_2)$, and not the readout of the ``wall clock'' in the laboratory~\cite{CuKu93}.
This idea was recently applied  to interpret
multi-speckle dynamic light-scattering data on an aging molecular glass former~\cite{Dyre22,Dyre23}. 
The actual parametrization 
is determined by the matching between the short time-difference behavior in the fast regime 
(which acts as a selection operator) and the long time-difference slow one. The analytic treatment of this matching is still open.

Importantly enough, the invariance of the Schwinger-Dyson equation 
needs a finite $T_{\rm eff}$ in the slow relaxation.
In the $p=2$ model, $T_{\rm eff}\to\infty$ and the symmetry is reduced to  {\it scale invariance}, $t\mapsto a t$~\cite{ChCuYo06}.
This is another distinction between the two types of models.
%The multi-thermalization organization of the slow relaxation of glassy systems, 
%the fact that all (interacting) observables evolving in the 
%same (two-) time scales share the same $T_{\rm eff}$ value~\cite{CuKuPe97,CuKu99},
%is possible thanks to the time-reparametrization~\cite{Kurchan22}. 
%Two systems brought into contact  should be able to rearrange their timescales 
%in order that all their $T_{\rm eff}$ match~\cite{CuKuPe97} and
%this time reorganisation is possible because the systems have a set of time-reparametrization
%invariances.

%We argued that the time reparametrization can be easily changed with a weak perturbation, 
%%as happens with non-potential or time-dependent forces as explained in Sec.~\ref{sec:driven-dynamics}.
%Instead,  certain relations between observables, like the fluctuation-dissipation ones, are 
%protected by the very same symmetry. 

It  was argued that time reparametrization invariance should be at the origin of 
dynamic fluctuations in glassy systems~\cite{Chamon02,Castillo02,ChCu07}. 
The fluctuations would then be due to dynamic fluctuations in the local time parametrizations
${\rm h}(\vec r, t)$ in a finite dimensional  system~\cite{Chamon02,Castillo02}. An effective theory for these fluctuations was constructed
and numerical and experimental tests were proposed and performed~\cite{ChCu07}. In the fully-connected models, similar fluctuations 
in the global observables are triggered by thermal noise in finite size systems. A sketch of how they would appear in a $\tilde \chi(\tilde C)$ relation, 
with $\tilde \chi$ and $\tilde C$ not averaged over noise is displayed in Fig.~\ref{fig:parametric}(c).
%, as found in the disordered models that occupy us here, see Equation~\ref{eq:qFDT}.

Such invariances have recently played a crucial role in the interpretation of the Sachdev-Ye-Kitaev (SYK) 
model~\cite{Sachdev} as a toy model of holography~\cite{Kitaev1,Maldacena}, see 
Sec.~\ref{sec:quantum}. 

\subsection{Higher order correlations  and length-scales}

The correlation $C(t_1,t_2)$ is playing the r\^ole of the order parameter of the dynamic transition
in the RFOT scenario as realized, e.g.,  by the spherical  $p\geq 3$ models.
Although there is no proper length-scale in this model, by analogy, the divergence of its fluctuations~\cite{KiTh88,Dasgupta91,Kob97,Biroli04},
\begin{equation}
G_4(t_1,t_2) = 
N^{-1} [ \langle ( \sum_i x_i(t_1)x_i(t_2) )^2  \; \rangle - \langle \sum_i x_i(t_1)x_i(t_2) \rangle^2 \; ]
\; , 
\label{eq:G4}
\end{equation}
has been used as an indication that a dynamic length 
scale diverges as the glassy arrest is approached from above. 
For vanishing $\epsilon = (T-T_d)/T_d$, the distance from the critical point, and in the
long time delay limit, $G_4$  becomes stationary and scales as
\begin{eqnarray}
G_4(t_1-t_2) = 
\left\{
\begin{array}{ll}
\dfrac{1}{\sqrt{\epsilon}} \; f_\beta\left( (t_1-t_2) \epsilon^{1/(2a)}\right) 
&
\qquad
t_1 -t_2 \sim \tau_\beta \sim \epsilon^{-1/(2a)} 
\; , 
\\
[\bigskipamount]
\dfrac{1}{\epsilon} \; f_\alpha \left( (t_1-t_2) \epsilon^{\gamma}\right)
&
\qquad
t_1 -t_2 \sim \tau_\alpha 
\; ,
\end{array}
\right.
\label{eq:scaling-G4}
\end{eqnarray}
%\item[(ii)]
in the $\beta$ and $\alpha$ regimes, respectively.
The scaling functions  satisfy
$f_\beta(x) \propto x^a$ for $x \ll 1$ and 
$f_\beta(x) \propto x^b$ for $x \gg 1$, while 
$f_\alpha(x) \propto x^b$ when $x \ll 1$ and 
vanishes at large $x$. Therefore, it progressively diverges upon $T$ getting closer to $T_d$ from above.
At fixed temperature below $T_d$, the role played by the distance from criticality $\epsilon$ is now 
the one of the waiting time. The scaling forms in  Equation~\ref{eq:scaling-G4} have to be modified accordingly~\cite{CoCuYo10}.

Experimentally, susceptibilities are easier to access than correlations.
Connections between 
$G_4$ and non-linear susceptibility via generalizations of the FDT, 
much in the same way as the ordinary correlation function is linked to the linear susceptibility,  
 have been established~\cite{Biroli04,LiCoSaZa08a,LiCoSaZa08b}.
This idea has been  investigated, see~\cite{book-heterogeneities} 
for an extensive account.
%The divergence of the susceptibility $\chi_4(t_1-t_2)$ was interpreted as a symptom of
%a dynamic growing length in the finite dimensional counterparts of these glassy models when approaching 
%$T_d$ from above. 
Attempts to go beyond this description, close to $T_d$ and in the $\beta$ relaxation regime, 
and write an effective theory of fluctuations were carried out in~\cite{Rizzo14,Rizzo16}, 
see also~\cite{Folena22}.
 
 \subsection{Large dimensionality}
 
 The equations discussed so far make no reference to real space; 
 they hold for systems with all-to-all interactions. 
 However, in physical systems interactions have a finite range, and one 
 should include this  ingredient in their description.

 In strongly interacting cases,  large dimensional expansions are a possible line of attack. 
%A first step towards the theoretical description of 
%the thermodynamics and dynamics of finite density particle systems in finite dimensions is to consider 
%systems of particles interacting through spherical pair potentials in spaces with large, actually diverging, dimensions.
The Langevin equation for an effective degree of freedom moving in an effective potential
%and in the presence of a colored noise, whose memory kernel  is determined self-consistently as the average of a force-force correlation, 
was derived in~Refs.~\cite{MaKuZa16a,MaKuZa16b}, in the spirit of the 
single variable Equation~\ref{eq:DMFT}. The key difference is that in the large $d$ treatment 
the memory kernel and effective noise correlation have to be determined by a self-consistent implicit functional of $C$ and $R$, 
while in Equation~\ref{eq:Sigma-D} -- corresponding to the schematic MCT -- they were simple functions of them.

%Since the structure of the single variable equations is so similar to the one of the pure spherical $p$ model, their solutions as well. 
The $d\to\infty$ equations under equilibrium conditions were
successfully analyzed both analytically and numerically. 
The ensuing qualitative features are the ones of the spherical $p$ models
but the details  are different (like the values of the exponents $a$ and $b$). 
Numerous facts of equilibrium hard sphere super-cooled liquids were derived with this
formalism, and numerical simulations confirmed the dimensional robustness of some of the predictions.
In particular, the critical properties of the jamming transition at infinite pressure were 
very successfully described by treating them as mechanically marginally stable packings~\cite{Charbonneau-etal}.
The out of equilibrium (aging) case has, for the moment, 
defied numerical integration and is not under control yet~\cite{Agoritsas21,Manacorda22}. A recent review~\cite{Charbonneau-etal} and a book~\cite{larged-book} 
summarized the technical aspects and outcome of the $d\to\infty$ method.

The dynamical transition in the $d\to\infty$ limit should be quite fragile and disappear in finite dimension. 
Activated processes should overcome the finite barriers between metastable states. 
The $d\to\infty$ approach could open the door to a systematic investigation of $1/d$ corrections. 

\subsection{Finite number of degrees of freedom}

 The simplest playground to gain insight on how to characterize these activated processes in a 
 large dimensionality space is to study the same models with large but finite number of degrees of 
 freedom.  Since activated processes should  take
an exponentially long time, $t \sim e^{AN}$, to complete, using finite and not too large $N$ may render them 
accessible numerically.
  
 Models with two-body interactions are especially simple in this respect. 
 The crossover between the dynamics 
 as in the $N\to\infty$ limit and the ones that feel the finite $N$ is
 controlled by the distribution of eigenvalues of ${\mathbb J}$ close to 
 its edge~\cite{FyPeSh,Barbier21}. 
 The low temperature relaxation from random initial conditions takes place in three regimes. The algebraic 
 time-dependencies in the first one are essentially the ones of $N\to\infty$ 
 Next comes a faster algebraic regime determined by the distribution 
 of the gap between the two extreme eigenvalues of ${\mathbb J}$, $\lambda_N-\lambda_{N-1}$.  Finally, an exponential regime takes over and it is 
 determined by the minimal gap sampled in the disorder average. Concerning  initial states which are almost projected on the saddles of the potential energy landscape, with
 deviations ${\mathcal O}(N^{-\nu})$ 
from perfect alignment,  the system escapes the initial configuration in a time-scale scaling as 
$\ln N$ after which the dynamics no longer “self-averages” with respect to the initial conditions. 

In the more interesting $p\geq 3$ cases, this analysis was mostly done with Monte Carlo simulations 
of the bi-valued (Ising) version~\cite{CrRi00a,Billoire05,Baity-Jesi18a,Baity-Jesi18b,Stariolo19,Stariolo20, Carbone22}. In particular,  the
 statistics of trapping times in the metastable states, trap energies, and energy barriers were considered 
 in~\cite{Stariolo19,Stariolo20,Carbone20,Carbone22}. One of the aims of these works was to infer, from these measurements
 which would be the best {\it trap model}~\cite{Bouchaud92,BouchaudDean1995,Barrat1995,Marinari18} description of the data.
 The program is not finished and more refined simulations and analysis are needed to reach a faithful conclusion.
 %These are only some initial steps in this  direction.

\subsection{Mathematical formalization}

%The analysis of  high dimensional landscapes, the derivation of the  Schwinger-Dyson equations, and the 
%study of their solutions have attracted the attention of probabilists and mathematical physicists. 
The literature on the formalization of the theoretical physicists derivations  and results is continuously growing. Some 
examples are the rigorous derivation of the Schwinger-Dyson equations~\cite{BenArous06}, the proof that the 
$p\geq 3$ model ages (on time scales that scale with $N$)~\cite{BenArous08}, and many refined analyses 
of the free-energy landscapes~\cite{Auffinger,Baik21}. 
    
\section{HAMILTONIAN DYNAMICS}
\label{sec:hamilton-dynamics}

The search for a statistical description of the asymptotic evolution of 
{\it isolated many-body quantum} systems has received much attention in recent years
due, in particular,  to the practical realization of closed ultra-cold atomic systems~\cite{Bloch08}, 
and the study of their evolution under various circumstances.
The main question posed in this context is: under which conditions a large closed system can act as a bath on itself and 
let local observables be described with a static average over a canonical Gibbs-Boltzmann
density operator? In cases in which this is not possible, the next issue is whether there is a density 
operator playing this role and if so which one.

These very same questions can be posed in a classical setting, whereby the classical mechanics evolution 
of an isolated many-body model, starting from selected initial conditions, is examined. 
Whenever  {\it ergodicity} applies,  time-averages and statistical averages of non-pathological 
phase space $(\vec p, \vec x)$
dependent macroscopic observables, $A(\vec p, \vec x)$, coincide.  More precisely, 
in the $N\to\infty$ limit there is a time $t_{\rm  erg}$ after which the time-average
\begin{marginnote}[] 
\entry{Ergodicity}{the equivalence of time and statistical averages of non-pathological observables.}
\end{marginnote}
\begin{eqnarray}
\overline{A(\vec p, \vec x)} \equiv \lim_{\tau\to\infty} \frac{1}{\tau} \int_{t}^{t+\tau} \!\! dt' \; A(\vec p(t'),\vec x(t'))
\end{eqnarray}
with $t>t_{\rm erg}$ reaches a constant which can also be calculated with the statistical average
\begin{eqnarray}
\langle A(\vec p,\vec x) \rangle \equiv \int d\vec p \, d\vec x \; \rho(\vec p,\vec x) A(\vec p,\vec x)
\end{eqnarray}
with a suitable $\rho(\vec p,\vec x)$. In standard statistical physics  the measure 
$\rho(\vec p,\vec x) $ is the microcanonical one, flat over the constant energy surface in phase space, 
when just the energy of the system is conserved. Alternatively, if one 
focuses on a selected part of the system, the measure becomes the canonical one, ${\mathcal Z}^{-1} \, e^{-\beta {\mathcal H}(\vec p, \vec x)}$, 
with an inverse temperature $\beta$ and the Hamiltonian ${\mathcal H}$ evaluated in the selected part
of phase space.  The inverse temperature fixes  $\langle {\mathcal H} \rangle$  where the average is taken over
the measure $\rho$, and controls the energy fluctuations.
If the system has a few other constants of motion, $I_k(\vec p, \vec x)$, the 
microcanonical measure is still flat over the corresponding reduced sector of phase space, 
and the canonical one is
%\begin{equation}
$
\rho(\vec p, \vec x) = {\mathcal Z}^{-1} \, e^{-\sum_k \gamma_k I_k(\vec p, \vec x)}
$
%\end{equation}
with $\gamma_k$ fixing the average  and the fluctuations of $I_k$.

In terms of  Equations~\ref{eq:defining-eqs}, 
the dynamics we are thinking about here corresponds to $\eta=0$, $\vec h=\vec 0$ and $f_i = -\partial V/\partial x_i$
with the $p$ body $V$.
Different kinds of initial conditions, which we called 
extended and condensed in Sec.~\ref{sec:rugged-free-energies}, basically correspond to the 
initial position of the particle being roughly anywhere on the sphere (extended) or localized 
close to the minima of the potential energy, as illustrated in Fig.~\ref{fig:sketch-states}(a) and (b)-(c), respectively. 
The equivalent of a {\it quantum quench} is to evolve with ${\mathcal H}$ a set of initial conditions prepared 
in thermal equilibrium with another ${\mathcal H}_0$. 
\begin{marginnote}[]
\entry{Quantum quench}{Usually performed quantum mechanically, the sudden change of one (or more) parameter(s) in the Hamiltonian of an isolated system.} 
\end{marginnote}
The simplest such quench is to rescale all the couplings  
\begin{equation}
J^{(0)}_{i_1, \dots, i_p} \mapsto J_{i_1, \dots, i_p} = \frac{J}{J_0} \, J^{(0)}_{i_1, \dots, i_p} 
\; , 
\end{equation} 
and  either inject ($J<J_0$) or extract ($J>J_0$)
a macroscopic amount of energy. 
%Different asymptotic behaviours, and their statistical descriptions, 
%can then be organized in a planar (energy variation  -- initial conditions)
%dynamic phase diagram, as we explain in the rest of this Section.
The problem at hand fits into the general scheme of the Schwinger-Dyson equations~\ref{eq:dyn-eqs-R}-\ref{eq:dyn-eqs-z}. 
with the terms proportional to $\beta_0$ replaced by 
\begin{equation}
\beta_0 \, (J_0/J)  \, D(t_1,0)C(t_2,0) 
\end{equation}
where $D =  J^2 (p/2) \, C^{p-1}$.
%As in the relaxational problem, one gets insight from the numerical solution of these equations and uses it to derive analytic 
%results in the long-times limit.
Not surprisingly, the outcome of such an experiment can be quite different depending 
on the type of $f_i$ used. 
Two classes of many-body systems can be distinguished, {\it integrable} and {\it non-integrable}, 
and two representatives instances of these are
the $p = 2$ and the $p=3$ disordered potentials, with the spherical constraint, respectively. 
The former turns out to be equivalent to Neumann's model of a particle confined to move on 
a sphere under the effect of anisotropic harmonic potentials~\cite{Neumann}.

\subsection{The non-integrable three-body ($p=3$) case}

\begin{figure}[h!]
%\hspace{3cm} (a) \hspace{3.5cm} (b) \\
\centerline{
\includegraphics[scale=0.24]{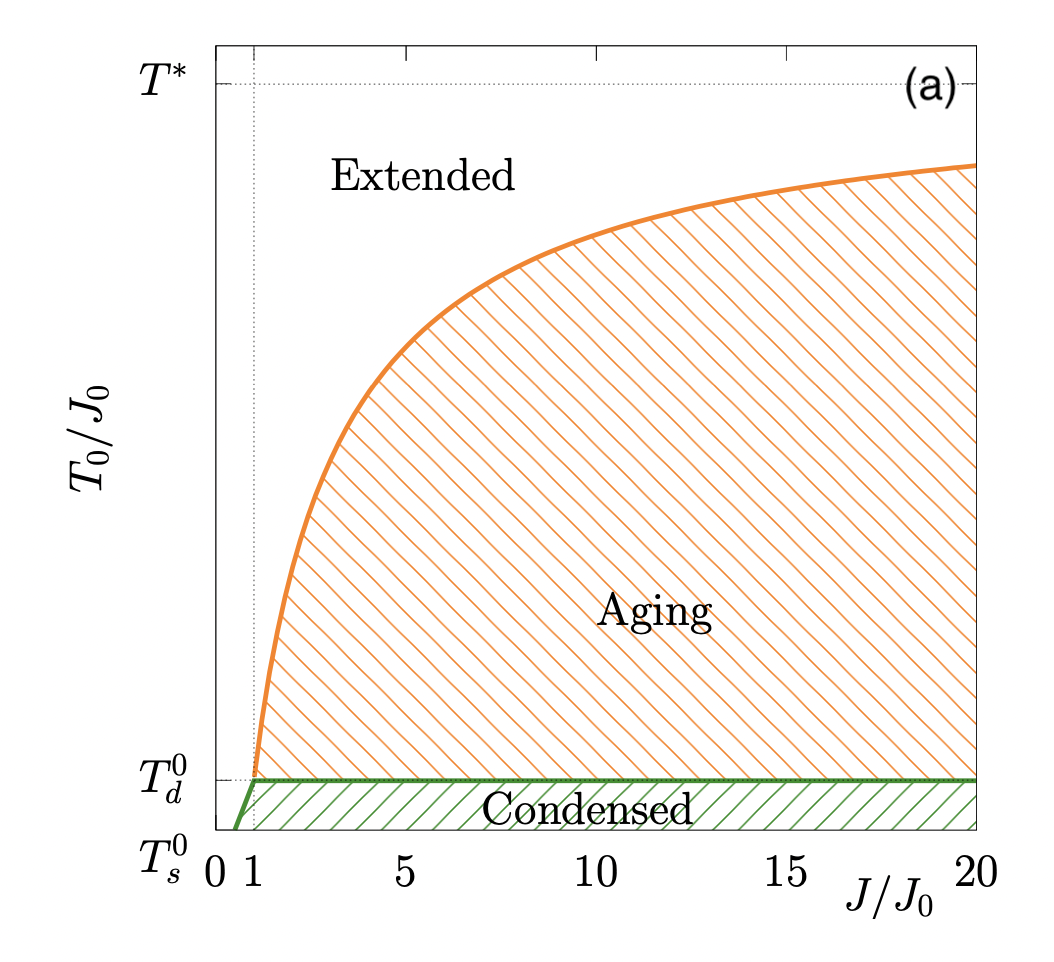}
\hspace{-9cm}
\raisebox{0.22cm}{
\includegraphics[scale=0.38]{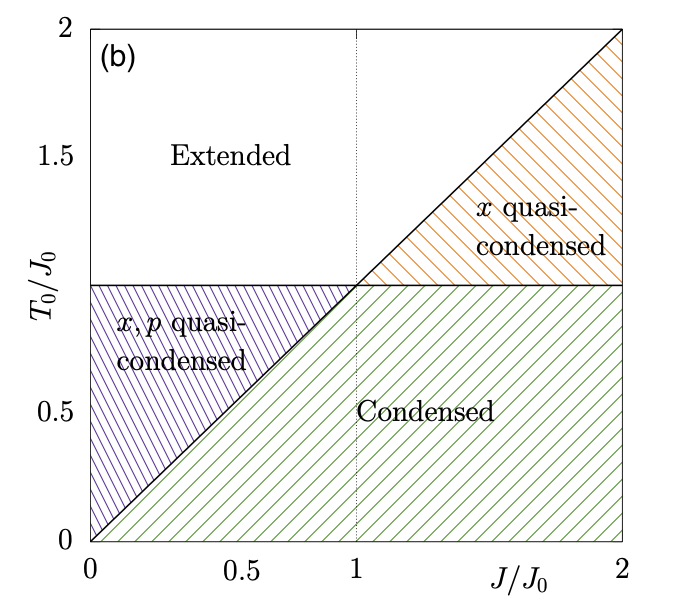}
}
}
\vspace{-0.01cm}
\caption{The dynamic phase diagram of the $p=3$ (a) and $p=2$ (b) isolated models. 
The initial conditions are drawn from the 
equilibrium Gibbs-Boltzmann distribution of the Hamiltonian ${\mathcal H}_0$
at inverse temperature $T_0$. The Hamiltonian evolution is performed with ${\mathcal H}$, 
differing from $H_0$ in a global rescaling of all couplings. The ratio $J/J_0$ measures
the injection ($<1$) or extraction ($>1$) of energy at the quench.
}
\label{fig:phase-diagram}
\end{figure}

In this case the dynamic phase diagram has three phases~\cite{CuLoNe17}
pictured in Fig.~\ref{fig:phase-diagram}(a). 
The statistical properties of the
long-term Hamiltonian dynamics of the $p\geq 3$ system are very similar to the relaxational
ones of the model in contact with a bath. In the Extended case, 
the particle has sufficient high energy to explore its available
phase space, it equilibrates to a conventional distribution and ergodicity at a $T_f$ which depends on $e_f$, the total energy, 
applies. In the Condensed case, the particle  remains
confined in the  well it was initiated in, and there is restricted Gibbs-Boltzmann equilibrium at temperature $T_f$. Finally, in the last Aging case 
one sets the system on the marginally stable threshold, 
a region in phase space in which the potential energy is dominated by
saddles points. The dynamics approach a non-stationary aging situation described by two temperatures, 
$T_f$ and $T_{\rm eff}$, that are also related to the final
energy of the system $e_f$ and other non-trivial properties of the quench. 
The details of the time dependent evolution 
are, of course, different from the purely relaxational case. The correlation and linear 
responses show oscillations. However, in the aging case, the oscillations 
exist only during the approach to the plateau, which also exists here, and 
are washed out in the slow regime where the relaxation becomes monotonic.

\subsection{The integrable two-body ($p=2$) case}
\label{subsec:Neumann}

This problem is almost quadratic and admits a very useful $N$ mode representation in which 
one focuses on the projection of the position and velocity vectors on  the 
eigenvectors $\vec v_\mu$ of the interaction matrix ${\mathbb J}$: 
$x_\mu(t) = \vec x(t) \cdot \vec v_\mu$, $\dot x_\mu(t) = 
\frac{d{\vec x}(t)}{dt} \cdot \vec v_\mu$, with $\mu = 1, \dots, N$.
In condensed initial conditions at $T_0<T^{(0)}_s=J_0$, $\langle x_{\mu=N}\rangle_{i.c.} = (qN)^{1/2}$ 
and all other $\langle x_{\mu(\neq N)} \rangle_{i.c.} = {\mathcal O}(1)$,  while in 
extended ones at $T_0>T^{(0)}_s=J_0$ also $\langle x_{\mu=N}\rangle_{i.c.}$ is ${\mathcal O}(1)$. 
%The ratio $J/J_0$ compared to one controls the amount of energy injected or extracted 
% in the instantaneous quench. 

This problem can be solved in two ways. First, one can simply integrate the Schwinger-Dyson Equations~\ref{eq:dyn-eqs-R}-\ref{eq:dyn-eqs-z} 
with the kernels in Equations~\ref{eq:Sigma-D}, $\eta=h=0$, 
$\gamma=1$ and $\overline V(q)=J^2 q^2/2$. In this calculation, the average over
the disorder ${\mathbb J}$ is done from the start. Second, one can proceed in a 
``mode resolved" and disorder dependent way.  $N$ dynamic equations for the modes $x_\mu$ and $\dot x_\mu$, 
coupled via the spherical constraint, can be efficiently studied numerically.
The most relevant projection is the one on the eigenvector associated to the eigenvalue at the edge of the spectrum, 
the $\mu=N$ one. 

The dynamic phase diagram extracted from the combination of these two kinds of studies 
is depicted in Fig.~\ref{fig:phase-diagram}(b).
It is characterized by the Lagrange multiplier (which is simply related to the difference between the 
kinetic and potential energies)
$z_f = \lim_{t\to\infty} z(t)$, the static susceptibility $\chi_{\rm st} = \lim_{t\to\infty} \chi(t_1,0)= \lim_{t\to\infty} \int_0^{t_1} dt \, R(t_1,t)$, 
the long time limit of the correlation with the initial condition
$\lim_{t\to\infty} C(t,0)$, and the occupation of the largest mode as measured by 
$\langle x_{\mu=N}^2\rangle_{i.c.}$.
From the analysis of  the Schwinger-Dyson equations and the mode resolved 
dynamics we distinguished four phases:
%depending on the static susceptibility $\chi_{\rm st}$, the long time limit of $C(t,0)$, and $\langle x_{\mu=N}^2\rangle$.
%They are the following.
\begin{enumerate}
%\item[(i)]
\item
 For $T_0>T_s^{(0)}$ and $T_0>J$  an extended phase  with $z_f = T_0 + J^2/T_0$, $\chi_{\rm st} = 1/T_0$, 
$ C(t,0) \to 0$, $\langle x_{\mu=N}\rangle_{i.c.} =\langle p_{\mu=N}\rangle_{i.c.} =0$, 
and $\langle x_{\mu=N}^2\rangle_{i.c.} = {\mathcal O}(1)$.
%\item[(ii)] 
\item
For $T_0>T_s^{(0)}$ and $T_0<J$ an extended phase  with $z_f =  2J$, $\chi_{\rm st} = 1/J$, $C(t,0) \to 0$, 
and quasi condensation of $\langle x_{\mu=N}^2\rangle_{i.c.} = {\mathcal O}(N^{1/2})$.
%\item[(iii)] 
\item
For  $T_0<T_s^{(0)}$ initial conditions aligned along the $N$th eigenvector of the matrix ${\mathbb J}$
and $T_0<J$, one finds a condensed phase with $z_f  =2J$, $\chi_{\rm st} = 1/J$, $C(t,0) >0$, 
and $\langle x_{\mu=N}^2\rangle_{i.c.} = q N$ with $q = {\mathcal O}(1)$.
%\item[(iv)] 
\item
For $T_0<T_s^{(0)}$ initial condition aligned along the $N$th eigenvector of the matrix ${\mathbb J}$
and $T_0>J$ a quasi-condensed phase of both $\vec x$ and $\vec p$
 with $z_f =T_0 + J^2/T_0 $, $\chi_{\rm st} = 1/T_0$, $C(t,0) \to 0$, 
and $\langle x_{\mu=N}^2\rangle_{i.c.} = {\mathcal O}(N^a)$ with $a<1$.
\end{enumerate}
%These features are summarized in the phase diagram in Fig.~\ref{fig:phase-diagram}(b).
In phases 1. (extended), 2. (quasi condensed in $\vec x$)
and 4. ($x$ and $p$ quasi condensed) a typical trajectory moves on the sphere and does not have
a macroscopic projection on any of the Cartesian axes given by the 
eigenvectors of the interaction matrix ${\mathbb J}$, see Fig.~\ref{fig:sketch-states}(b). 
In phase 3., a typical trajectory starting from a symmetry broken initial condition
with a macroscopic projection on the direction of the $N$th coordinate keeps this 
projection in the course of time and, typically, precedes around it. 
\comments{
The GGE as well as the dynamic averages of the momentum in
this same $N$th direction, $\langle p_{\mu=N}\rangle_{\rm GGE} = 
\overline{\langle p_{\mu=N}\rangle_{i.c.}}=0$ while 
$\langle x_{\mu=N}\rangle_{\rm GGE}  = 
\overline{\langle x_{\mu=N}\rangle_{i.c.}}={\mathcal O}(N^{1/2})$.
}
In none of these phases the quasi quadratic Neumann model equilibrates 
to a Gibbs-Boltzmann measure. Accordingly, there is no
single temperature characterizing the values taken by different observables in the long time limits, not even
after being averaged over long time intervals. Instead, 
in the infinite size limit, and after the Lagrange multiplier saturates to its asymptotic value
$z_f$, 
the modes decouple, and each of them behaves as if in equilibrium at its own 
temperature $T_\mu$:
\begin{equation}
\overline{\langle p_\mu^2\rangle}_{i.c.} = k_BT_\mu
\; , 
\qquad\qquad\quad
\overline{\langle x_\mu^2\rangle}_{i.c.} = \frac{k_BT_\mu}{z_f -\lambda_\mu} 
\; . 
\end{equation}
 %There is therefore a non-trivial spectrum of mode temperatures $\{T_\mu\}$. 
The functional dependence of $T_\mu$ on the parameters  is different in each phase. 
The co-existence of modes at different temperature is possible since they get 
effectively decoupled asymptotically. 

Interestingly enough, the connection with the idea of measuring a time-delayed (or frequency) dependent effective
temperature $T_{\rm eff}$ from the deviations from the fluctuation-dissipation relation of the global 
correlation, $C(t_1, t_2) = N^{-1} \sum_\mu [\langle x_\mu(t_1) x_\mu(t_2)\rangle_{i.c.}]$ and the corresponding linear response, 
 then establishes very naturally~\cite{FoGaKoCu16,deNardis17}. 
 Each oscillator has its own frequency $\omega^2_\mu = (z_f-\lambda_\mu )/m$.
The global frequency dependent effective temperature is defined by $k_B \overline T_{\rm eff}(\omega) \equiv 
- \omega \tilde C_{\rm st}(\omega)/{\rm Im} \tilde R_{\rm st}(\omega)$, where the 
tildes indicate Fourier transform with respect to time delay. Setting the measuring frequency 
$\omega$ to $\omega_\mu$ one selects a particular mode 
and hence measures its own temperature, $\overline T_{\rm eff}(\omega=\omega_\mu)=T_\mu$.

 In the long-time limit the temporal averages coincide with the statistical ones calculated with a 
Generalized Gibbs Ensemble (GGE), 
\begin{marginnote}[]
\entry{Generalized Gibbs Ensemble}{An extension of the canonical measure that includes all constants of motion in
the exponential Boltzmann factor.}
\end{marginnote}
\begin{equation}
\rho(\{\dot x_\mu, x_\mu\}) = {\mathcal Z}^{-1}(\{\gamma_\mu\}) \;\; 
e^{-\sum_{\mu=1}^N \gamma_\mu I_\mu(\{\dot x_\mu, x_\mu\}) }
\; . 
\end{equation}
The $I_\mu$ are the $N$ integrals of motion in involution, quartic functions 
of the phase space coordinates $\{ x_\mu, \dot x_\mu \}$~\cite{Uhlenbeck}
parametrized by the  post-quench 
eigenvalues $\lambda_\mu$ of ${\mathbb J}$:
\begin{equation}
I_\mu(\{\dot x_\mu, x_\mu\}) = x_\mu^2 + \frac{m}{N}
\sum_{\nu(\neq \mu)} \frac{(x_\mu {\dot x}_\nu- x_\nu {\dot x}_\mu)^2}{\lambda_\mu-\lambda_\nu}
\; , 
\end{equation}
 The $\gamma_\mu$ are Lagrange multipliers which are fixed by the requirement, 
\begin{equation}
\langle I_\mu\rangle_{\rm GGE} = I_\mu(0)
\; , 
\end{equation}
with $I_\mu(0)$ the 
initial values right after the quench, which in any case are conserved by the dynamics.
One then successfully verifies that, in all phases, 
\begin{equation}
\overline{\langle p_\mu^2\rangle}_{i.c.} = \langle p_\mu^2\rangle_{\rm GGE} 
\; , 
\qquad\qquad\quad
\overline{\langle x_\mu^2\rangle}_{i.c.} =  \langle x_\mu^2\rangle_{\rm GGE} 
\; . 
\end{equation}

This is a particularly exciting integrable model for which one can tune the initial conditions to have 
radically different properties and also induce different  kinds of phase transition with the quenches. 
The fact that the GGE describes its asymptotic dynamics was not granted {\it a priori}. 
%The model could be solved in depth with the Schwinger-Dyson method, and one could also get more detailed information 
%of the mode behavior thanks to its almost quadratic character. 
The knowledge we had of its canonical equilibrium 
behaviour and relaxational dynamics were valuable guidelines to find the solution under Newton dynamics.
It is to stress, though, that this  is not 
a quadratic model because of the spherical constraint, and it is for this reason that it can 
support a non-trivial phase diagram.

%\section{NON-RECIPROCAL INTERACTIONS}

\section{DRIVEN DYNAMICS}
\label{sec:driven-dynamics}

Non-reciprocal interactions
violate detail balance and inhibit equilibration. Specially interesting evolutions 
arise in glassy  systems  in the potential case and subjected to non-potential forces. 
On top of the possible 
currents generated, new hallmarks such as 
large number of attractors in the form of 
fixed points, limit cycles and chaotic evolutions may also exist. This Section deals with systems 
driven in different ways.

\subsection{Rheology and active matter}

Shearing, due to a non-conservative force applied on the 
boundaries of a system,  is a common way of favoring the 
relaxation of glassy or jammed materials ({\it shear thinning}).
This mechanical perturbation, which exerts work,
can be mimicked with kernels $\Sigma$
and $D$, in Equations~\ref{eq:dyn-eqs-R}-\ref{eq:dyn-eqs-z}, that do not satisfy an FDT-like relation, that is
$\Sigma \neq D' R$.
%stemming from non-conservative forces that exert mechanical work. 
The relaxation time then decreases with the drive, $\tau_\alpha(\gamma)$, 
and this results in the suppression of the aging process at low temperatures. 
For forces of the $p$-body kind, such that in the absence of drive there is a 
glass transition mechanism,  $C$ and $R$ are stationary but
still display a two-step relaxation pattern, similar to the one observed at equilibrium slightly above $T_d$, if the 
perturbation is not too strong.
In the weak drive limit, $C(t_1, t_2) = C_{\rm rapid}(t_1-t_2) + C_{\rm slow}((t_1-t_2)/\tau_\alpha(\gamma))$, 
with limiting values as the ones in Equations~\ref{eq:limit-fast},~\ref{eq:limit-slow}, and
violations of FDT consistent with a 
two-temperature scenario~\cite{CuKuLePe97,Berthier00}. Interestingly, the Monte Carlo simulation of 
these models with finite number of variables allows one to see jumps between the different metastable 
states that have long but finite lifetime. The non-potential forces let the system transit more easily over 
barriers and explore the landscape faster.  Building upon the large $d$ formulation of 
the pure relaxational dynamics, Agoritsas {\it et al.} recently
derived  and studied the large $d$ equations that govern the dynamics of interacting particles under 
shear~\cite{Agoritsas19a,Agoritsas19b}. 

Time-dependent drives also pump energy into a system. 
An example, relevant to describe vibrated granular matter~\cite{BeCuIg01} but also much 
studied quantum mechanically, 
is the one of a global periodic force, such as $h(t) =h \sin(\omega t)$. If the system  is also connected to a thermal bath, it may 
reach a periodic ``Floquet'' regime, with monotonic  stroboscopic behavior. 
The stemming dynamic phase diagram has three axes, $T/J$, $h/J$ and $\omega t_0$, with the frequency $\omega$ 
of the drive being another control parameter.
For $p=2$ potentials, a critical line $h_c(\omega)$ going as $\omega^{1/4}$ for $\omega \sim 0$ at $T=0$, and increasing 
with $\omega$,  was estimated. Below this curve aging with superimposed oscillations, that are washed out when 
observing stroboscopically at times $t_n=2\pi/\omega$, survives.
In the $p=3$ case, the peculiar feature $\lim_{\omega\to 0} h_c(T/J, \omega)< h_c(T/J, \omega=0)$ was determined. 
The simulation of finite size systems with the periodic perturbation confirms the existence of metastable 
states that, however, do not block the dynamics completely. Jumps between these states are promoted by the injection of energy 
provided by the drive.

Yet another way to make a system time-translational invariant is to consider {\it annealed disorder}, that 
is change it slowly in time~\cite{Horner92} with some time-scale $\tau_{\rm ann}$, which sets the structural relaxation time, 
and $C(t_1,t_2) \sim C_{\rm slow}((t_1-t_2)/\tau_{\rm ann})$ for long time differences.
%We reckoned that a random force with no associated dissipation corresponds to a special 
%thermal bath with finite friction and infinite temperature. 
One can also consider the coupling 
to several baths with different temperatures and characteristic times as a way of generating 
non-equilibrium dynamics~\cite{CuKu00}.
\begin{marginnote}[]
\entry{Annealed disorder}{It slowly changes in time.}
\end{marginnote}

Activity acts at the single particle level. 
A simple way to model it is to add random forces $h_i$ to the evolution equations, which are not 
accompanied by a corresponding dissipative term~\cite{Berthier13}. This is equivalent to adding 
$T\to \infty$ noise while keeping associated $\eta$ finite in Equations~\ref{eq:defining-eqs}. The ``bath'' is then made of two components: 
a normal one in equilibrium at $T<\infty$ and a non-equilibrium one which does not satisfies FDT and drives the system out of equilibrium. 
A choice has been to use 
independent random forces with zero mean and exponentially decaying temporal correlations
with a characteristic time-scale $\tau_{\rm act}$, which 
competes with the internal time-scale of the structural relaxation.
While the equilibrium glass transition disappears in 
the presence of power dissipation of infinitesimally small amplitude~\cite{CuKuLePe97}, 
it survives the introduction of these fluctuating forces, even of large amplitude. 
Time correlation functions display a two-step decay reminiscent
of the unforced behaviour. The location of the transition decreases for increasing 
driving. There is a non-equilibrium $T_{\rm eff}$  for the slow degrees of freedom even for the stationary fluid phase, 
not only deep into the glass~\cite{Berthier13}. The large $d$ description of active systems has also been addressed~\cite{Arnoulx19,Arnoulx21}.

%Heading 1
\subsection{Theoretical ecology}
\label{subsec:ecology}

In theoretical ecology, coupled differential equations
rule the time evolution of the population densities of different species
%, $x_i$, 
in interaction which are, possibly, also coupled to an environment that can bring noise and 
immigration into the community. 
%Much work has been devoted to the analysis of such models with a few species. 
Recently, focus has turned to the analysis of ecosystems with large number of species, 
a limit in which statistical physics methods can be applied. 
Concrete examples of $N\gg 1$ communities range from  microbes  in  the  gut  to  plants  in  a  rain 
forest. In their simplest modeling  there
is no spatial structure, an assumption justified for  
%: all individuals from all species can interact in a way that does not depend on the spatial location. Therefore, it
well-mixed ecosystems. 
The interactions are particularly difficult to infer in diversity-rich cases
and are usually taken to be controlled by quenched random parameters.
%This modelling is also deterministic, but it has a probabilistic component as we will specify below.
The similarity with disordered mean-field physical models then becomes apparent.

The population sizes $x_i$ are non-negative real variables that must remain so throughout the dynamics. 
If the {\it absorbing} value $x_i = 0$ is reached at a given time, the 
population goes extinct and so remains 
subsequently if there is no immigration from the external world. A broad class of differential equations of the form $dx_i/dt = 
g_i(\vec x)  x_i$, with $\vec x = (x_1, \dots, x_N)$ and  initial values $x_i(0) >0$   satisfy these conditions. 
\begin{marginnote}[]
\entry{Absorbing value}{Once reached it cannot be left.}
\end{marginnote}

\subsubsection{Linear stability}

A fundamental question in ecology is whether very diverse
systems made of a large variety of interacting species 
are more resilient,
%, and recover better from perturbations displacing 
%them from  (multidimensional) {\it fixed points}, which with a little abuse of notation are also called equilibria,
or more unstable, than small size ones. 
The stability of such large complex systems can be 
quite generically studied  using a linear set of equations, expected to represent 
the dynamics close to a {\it fixed point}, also named {\it equilibrium} in this context. Calling $y_i = x_i-x_i^{*}$ the 
deviation of the population density of the $i$th species from its equilibrium value $x_i^{*}$,
the equations read 
\begin{eqnarray}
\frac{dy_i}{dt} = - \mu_i y_i +\sum_{k=1}^N J_{ik} y_k 
\label{eq:may}
\end{eqnarray}
for $i=1,\dots, N$, and $N$ the number of species. The parameters $\mu_i$ are positive and typically taken to be all equal to $\mu$. 
%Their positivity ensures that 
In the absence of interactions, $J_{ik}=0$ for all $i,k$, 
the system is self-regulating and any $y_i \neq 0$ returns to zero exponentially fast. 
%with the time-scale $1/\mu$.  
The  {\it community matrix}   ${\mathbb J}$ has elements $J_{ik}$ 
and measures the {\it per capita} effect of 
the $k$th species on the $i$th one  at the presumed equilibrium. Typically, it is a
non-symmetric, $J_{ik} \neq J_{ki}$, square matrix. 

%The interactions between species are particularly difficult to infer in diversity-rich ecosystems. 
Following previous numerical studies~\cite{Gardner}, May interpreted the $J_{ik}$ as the entries of a random matrix
${\mathbb J}$ with averaged number of non-vanishing elements,  the 
averaged connectivity of the ecological network, equal to $C>0$~\cite{May72}. 
If species $k$ has no effect on species $i$, $J_{ik} =0$. Otherwise, the non-zero 
elements are  independently sampled from a probability distribution with zero mean 
and variance $\sigma^2$. 
The assumption of there being as many positive as negative values of 
%the pairwise interactions 
$J_{ik}$ is a plausible one for large ecosystems. 
Stability 
%of the linear system in Equation~\ref{eq:may} 
 is ensured if 
all the eigenvalues of  ${\mathbb J}$ have real parts less than $\mu$. Random matrix theory then 
establishes that in the large 
$N$ limit
%the largest real part of the eigenvalues goes as $\sigma \sqrt{CN}$, and 
the system is almost certainly stable if $\mu/(\sigma \sqrt{CN}) >1 $  and 
 unstable if  $\mu/(\sigma \sqrt{CN}) <1 $, with a sharp transition between the two.
%  types of behavior upon  changing $\mu$, $\sigma$, or $\sqrt{CN}$. 
 For fixed $\mu$, the dynamics will almost certainly become unstable 
 for sufficiently large $\sigma \sqrt{CN}$, that is 
 % The main conclusion of May's analysis was that an ecological system can become unstable 
 for sufficiently large complexity as quantified by the connectivity and averaged interaction 
 strength.  The larger the system size, the more pronounced the effect, in the flavor of phase transitions.
 If, initially, the bound is violated, the ecosystem will drive some species to extinction 
 until a stable community with a number of species satisfying the bound remains. 
 
 This surprising result was questioned  since very diverse ecosystems, not satisfying this bound, 
 do exist in Nature~\cite{McCann00} (e.g. $N\sim 10^5$ for the oceanic plancton). 
 The model has  been enriched in many ways to avoid May's
 restrictive limit~\cite{Allesina12,Allesina16}. It is not the scope of this article to present a 
 comprehensive review of theoretical ecology, it just means to illustrate how the methods and 
 ideas of the non-equilibrium dynamics of disordered physical systems can be adapted to describe 
 problems in this area. 
 
A similar stability analysis can be applied to other areas, {\it e.g.} the functioning of 
neural networks~\cite{Sompolinsky88,Wainrib13}, 
systemic risk in trading~\cite{Farmer13}, or  large economies~\cite{Moran19}.
  
 \subsubsection{Instability: explosion of the number of fixed-points}
 
Following the studies of rugged free-energy landscapes recalled in Sec.~\ref{sec:rugged-free-energies} 
the natural project to carry out for dynamical systems with random non-conservative 
forces $f_i$ is the enumeration and stability analysis of the fixed points. 
Fyodorov and Khoruzhenko went back to the evolution of the 
relative population sizes $x_i$'s and studied the model~\cite{Fyodorov}
\begin{eqnarray}
\frac{dx_i }{dt} 
&=& 
- \mu x_i + f_i(\vec x) 
\qquad
\mbox{with}
\qquad
f_i(\vec x) 
= 
- \frac{\partial V(\vec x)}{\partial x_i} + \sum_{k} \frac{\partial A_{ik}(\vec x)}{\partial x_k}
\; .
\label{eq:Fyodorov}
\end{eqnarray}
Each species becomes extinct on its own while the interactions,
mimicked by the forces $f_i$, allow for  the persistence of the community.
The random forces are separated in potential and non-potential (divergence free) contributions.
$V$ and  $A_{ik} = -A_{ki}$ are statistically independent and Gaussian distributed, with
$[V]=[A_{ik}]=0$,
$[ V(\vec x)V(\vec y) ] = \nu^2 \, \overline V(|\vec x -\vec y|^2)$ and 
$[ A_{ij} (\vec x) A_{lm}(\vec y) ] = a^2 N \, \overline A(|\vec x -\vec y|^2) (\delta_{il}\delta_{jm} -\delta_{im}\delta_{jl})
$. 
%In the neighbourhood of an equilibrium state, $d\vec x/dt=0$, May's Equation~\ref{eq:may} is 
%recovered with $J_{ik} = \partial f_i/\partial x_k$ evaluated at $\vec x^*$. 
%This opens the possibility of parametrising the models by the statistics of the functions $f_i$. 

The authors  put the accent on the evaluation of the averaged number of zeroes of the total force, 
adapting the Kac-Rice method  for counting solutions of simultaneous equations~\cite{Kac43,Rice44}
and using random matrix theory (annealed calculation).
They found an abrupt transition at $\mu/(2\sigma\sqrt{N})=1$ with $\sigma=\sqrt{\nu^2+a^2}$
between a 
%{\it topologically trivial phase} 
a phase with, on average, a single equilibrium and a non-trivial one 
 with an averaged number of fixed points growing exponentially with 
 $N$~\cite{Fyodorov} (for $a=0$ see~\cite{Bray07}). 
 This transition should be  the reason for May's instability.
 The vast majority of fixed points are unstable beyond it
 and, as explained below, induce  long relaxation times and chaotic dynamics. At fixed $\nu^2/\sigma^2$, 
and decreasing $\mu/(2\sigma\sqrt{N})$  there is a new transition towards a phase with stable fixed points~\cite{BeFyKo21}.

%This scenario should be shared by other systems of randomly coupled autonomous 
%ordinary differential equations with large numbers of degrees of freedom.
A similar explosion of complexity was measured in a neural network model~\cite{Wainrib13}, with 
randomly interconnected neural units with $f_i = \sum_{k(\neq i)}J_{ik} S(x_k)$, $S$ an odd sigmoid function representing the synaptic non-linearity, 
and  $J_{ik}$ independent Gaussian variables, 
 representing the (non-symmetric) synaptic connectivity between neurons $i$ and $k$. 
 
\subsubsection{The random Lotka-Volterra model}

%Ecosystem can have rather complex evolutions which go beyond May's linear stability analysis.
%and not just admit weak perturbations around static situations. 
%Moreover, May's description does tells us 
%much about what happens in unstable situations either. 
%Going back to the description of the population densities $x_i$,  one postulates
A particular form of Equation~\ref{eq:Fyodorov} is
\begin{equation}
\frac{dx_i }{dt} = 
g_i(x_i) - x_i \sum_{k(\neq i)} \alpha_{ik} x_k  
%\sum_{k (\neq i)} f_{ik}(x_i,x_k)
+ \lambda_i
\; . 
\label{eq:LV}
\end{equation}
%The first term in the right-hand-side, 
$g_i(x_i)$, describes the autonomous  dynamics 
of species $i$ as if it were isolated from the rest, and decides  whether it survives  or disappears
on its own. In the former case, after some relaxation time $x_i \to x_i^*$, 
 the {\it carrying capacity} of the species. A common choice is 
% \begin{equation}
$g_i(x_i) = \frac{\mu_i}{K_i}  x_i (K_i-x_i) $
%\label{eq:g}
% \end{equation}
 with zeros at
$x^*_i = 0$ (extinction) and 
$x^*_i = K_i $ (saturation).
 The second term in the right-hand-side %of Equation~\ref{eq:LV} 
 models binary interactions, which could be 
 generalized to  include higher-body terms. 
 The signs of $\alpha_{ik}$ represent the effects of one species on the other. 
 If species $k$ is a predator of species $i$, 
 $\alpha_{ik}$ is positive and tends to reduce the population $x_i$. If the two species compete for the same resources, 
 both $\alpha_{ik}$ and $\alpha_{ki}$ are positive.  Both are negative if there is mutualism. $\lambda_i$ is an immigration rate
 which is usually taken to be uniform across species, $\lambda_i=\lambda$.
 In the spirit of May's approach, the coefficients $\alpha_{ik}$ are taken to be 
 i.i.d. random variables. In the simplest description all species are connected to 
 all others and the moments
 are $[\alpha_{ik}] = \alpha/N$, $[(\alpha_{ik}- [\alpha_{ik}])^2] = \sigma^2/N$
 and $[(\alpha_{ik}- [\alpha_{ik}])(\alpha_{ki}- [\alpha_{ki}])] = \gamma\sigma^2/N$.
 %, where the square brackets indicate an average over their probability distribution. 
 The scaling with $N$ ensures a proper large $N$ limit. The parameter
 $\gamma$ measures the asymmetry of the interactions; it  ranges from $-1$ 
 (fully antisymmetric, all interactions are of predation-prey type) 
 to $1$ (fully symmetric, an energy can be defined).
A  linear expansion around a putative fixed point $x_i^*$ yields May's Equation~\ref{eq:may}
with a simple relation between
$\mu$, ${\mathbb J}$, $g_i$ and the $\alpha_{ik}$.
Finally, one could add {\it demographic noise}
accounting for deaths, births, and other unpredictable events. It is usually chosen to be a 
Gaussian random variable with zero mean and 
\begin{equation}
\langle \xi_i(t_1) \xi_j(t_2) \rangle = 2 T \, x_i(t_1) \, \delta_{ij} \delta(t_1-t_2)
\; . 
\label{eq:multiplicative-noise}
\end{equation}
%The fact that the population $x_i$ appears in the amplitude makes the noise {\it multiplicative}.
The {\it multiplicative} form of Equations~\ref{eq:LV}-\ref{eq:multiplicative-noise}.
ensures that
%The population densities $x_i $ are non-negative variables that must remain so throughout the dynamics. 
for $\lambda_i=0$
if the absorbing value $x_i = 0$ is reached  the 
population goes extinct and $x_i$ remains zero at all later times. The  noise is treated in the Ito convention~\cite{Roy19,Kesslerand,Bunin17,Galla18,Pearce20,Altieri21}. 

The model has five control parameters: the average strength 
$\alpha$ and the variety $\sigma$ of the interactions, their asymmetry~$\gamma$, 
the immigration rate $\lambda$
and the temperature  $T$ of the demographic noise.
In the $N\to\infty$ limit, the single variable equations are~\cite{Roy19,Galla18}
\begin{equation}
\frac{dx(t_1)}{dt_1} = x(t_1) \left[1-x(t_1)-\alpha m(t_1) -\sigma \zeta(t_1) + \gamma \sigma^2 \int_0^{t_1} \!\!  dt \, R(t_1,t) x(t) + h(t) \right]
\label{eq:DMFT-rLV}
\end{equation}
with the averaged population, self correlation and linear response
\begin{equation}
m(t_1) = [\langle x(t_1)\rangle ]
\, , 
\quad
C(t_1,t_2) =[ \langle x(t_1)x(t_2)\rangle ]
\, ,
\quad
R(t_1,t_2) = [\langle \left. \frac{\delta x^{(h)}(t_1)}{\delta h(t_2)} \right|_{h=0} \!\!\!\! \rangle ]
\,  ,
\end{equation}
where the angular brackets denote average over initial conditions and the effective noise 
$ \zeta(t_1) $ with zero mean and self-consistent 
correlations $\langle \zeta(t_1) \zeta(t_2)\rangle=\langle x(t_1) x(t_2)\rangle$. For the sake of 
compactness we set the demographic noise and the immigration rate to zero and we 
absorbed $\mu_i$ and $K_i$ in a redefinition of the $x_i$. 
Importantly enough, the original system 
(with no demographic noise) was deterministic, but the effective single variable one 
is stochastic. 
The fully-connected model without demographic noise at fixed $\gamma$ 
has three phases~\cite{Bunin17,Roy19}:
\begin{enumerate}
%\item[(i)]
\item
{\it Single fixed point} for small $\sigma$. In the long time limit, all initial conditions reach the same 
saturation values, $x_i^*$ (some $x_i^*$ may vanish if the $i$th species goes extinct).
The $x_i^*$ are Gaussian distributed, truncated at negative values.
This state is stable against local and global perturbations. The global 
correlation functions are stationary. 
%\item[(ii)]
\item
{\it Multiple Attractors} for large $\sigma $ and small immigration rate $\lambda \stackrel{>}{\sim} 0$.
The single stable fixed point loses its stability at $\sigma_c= \sqrt{2}/(1+\gamma)$, and not only the {\it averaged}
but even the {\it typical} number of equilibria grows exponentially with the number of species~\cite{Roy22}. 
The system cannot settle in a fixed point and this phase exhibits {\it chaotic  dynamics}, with large fluctuations 
of the species populations, for $\gamma\neq 1$
or aging dynamics, as for physical systems, for $\gamma=1$. Chaos is self-destructive for $\lambda=0$. 
Similar features were observed in 
neural networks~\cite{Sompolinsky88}, evolutionary game theory~\cite{Opper92,Galla06}
and replicator equations with nearly antisymmetric random interactions~\cite{Pearce20}.
\begin{marginnote}[]
\entry{Chaotic dynamics}{The populations do not stabilize and undergo large fluctuations.}
\end{marginnote}
\item
{\it Multiple Attractors} for large $\sigma $ and vanishing immigration rate $\lambda=0$.
The system ages with a rather different mechanism from the one 
of the  physical $p$-body models which do not have absorbing boundary values. 
Here there are near-extinction processes, 
whereby some population sizes go very close to zero for some time before rebounding, 
accompanied by a global slowdown of the dynamics. Asymptotically, the time 
it takes for a variable to leave the vicinity
of an absorbing value to visit the vicinity of another is
proportional to the age of the system. This mechanism was coined {\it aging by near extinctions}~\cite{Arnoulx23a} 
and it is due to the fact that all the fixed points
are unstable~\cite{Ratzke20},  contrary to the marginal
threshold reached in usual glassy dynamics~\cite{CuKu93}.
\begin{marginnote}[]
\entry{Aging by near extinctions}{The progressive dynamic slowdown due to the approach to 
unstable fixed points.}
\end{marginnote}
 %\item[(iii)]
 \item
{\it Unbounded Growth} for large negative $\alpha$. In this case the interactions are so cooperative
that the dynamics can override the single-species saturation.
%, and a
 %subgroup of species grow without bound, while  all the others die out. 
 The unbounded growth
 is a pathology which could be cured by a stronger saturation  imposed by 
 $g_i(x_i)$. For instance, a cubic term of the form $g(x_i)= -\mu_i x_i (1 - x_i/K_i)(x_i-m_i)$
avoids the unbounded growth but 
it also drastically changes  the phase diagram~\cite{Altieri22}.
 \end{enumerate}
%\end{enumerate}
%The {\it typical} (quenched) number of fixed points was recently evaluated in~\cite{Roy22} and although 
%smaller than the averaged (annealed) one, there is an extended phase in which it is still exponential in $N$.
In these explicitly out of equilibrium dynamics the deviations from the standard FDT have not been 
characterized yet. It remains, certainly, as an interesting open problem that should be at reach 
of the numerical analysis.

Another issue concerns the effects of demographic noise~\cite{Altieri21}. 
At fixed $\gamma$, $\alpha$ and $\lambda$,  the phase diagram, 
parametrized by temperature and the species variability $\sigma$, has three phases. In the symmetric case $\gamma=1$, one 
phase has a single fixed point
in which the relaxation time is finite and the correlation decays as a function of time difference. Another one 
has multiple attractors and the relaxation ages, similarly to what is found in glassy systems. Finally, 
a phase in which the multi-equilibria are organized in a hierarchical manner ({\it a.k.a. Gardner phase})
exists at still lower temperatures.
When the asymmetry is considered, the dynamics in the multi-attractor  phases become chaotic 
%as for neural networks and spin-glass models in the presence of asymmetric couplings, 
and it has an indefinitely long lifetime for $\lambda>0$, 
while it is replaced by aging for $\lambda=0$.

\subsection{Optimization and learning}
\label{subsec:acceleration}

How to identify the absolute minimum of a function defined on a high dimensional space is 
the key issue in hard optimization problems. Usually,  the algorithms used to attain this goal are based on 
physical rules. Popular choices are Langevin dynamics satisfying detailed balance, or Monte Carlo rules also respecting
this condition. Such evolutions eventually minimize the 
energy or cost function of a finite size system if the zero temperature limit is taken in a convenient way. 
However, this strategy can be very inefficient for systems with complex free-energy landscapes, 
needing a number of operations that scale as
the exponential of system size. Alternatives have been searched for since long ago.
In {\it simulated annealing}, {\it cluster Monte Carlo methods}  and other,  
the physical restrictions are lifted. In this section we briefly mention some of the recent progress in this direction.
%, keeping close to the main subject of this article.

Langevin dynamics with divergence-less forces
which do not respect detailed balance, have been proposed to accelerate 
the diffusive dynamics while respecting the asymptotic 
statics measure~\cite{Hwang05,Lelievre13,Ichiki13}. 
While the dynamics pf $p$-body models are indeed accelerated, with times rescaled by a factor, 
the dynamic transition at $T_d$ is not modified~\cite{Ghimenti22}. Basically, there is an 
enhanced mobility which accelerates the evolution without affecting the free-energy 
landscape.

Understanding the extraordinary performance of 
{\it artificial neural networks}, in the deep learning or perceptron realization, is of extreme importance.
The {\it perceptron}~\cite{McCulloch43,Rosenblatt58} is a simple model of a natural or artificial neuron: 
\begin{marginnote}[]
\entry{Perceptron}{A model neuron.}
\end{marginnote}
it receives an input, and once multiplied by the learned weight coefficient, generates an output.  
In {\it machine learning}, networks of such perceptrons are used as classifiers.
The {\it feed-forward networks}, at the basis of modern {\it deep neural networks},
are made of successive layers of such neurons. 
\begin{marginnote}[]
\entry{Deep neural networks}{Feed-forward networks with many internal layers.}
\end{marginnote}
The perceptron problem  consists in,  given a set of inputs 
and another set of associated outputs, find the synaptic weights, connections and strengths, such that the 
network makes a correct classification. Training a network with {\it supervised learning} corresponds to
minimizing a cost function that depends on the parameters of the network and the dataset.
The latter can also be seen as a {\it constraint satisfaction} problem: 
\begin{marginnote}[]
\entry{Constrained satisfaction problems}{require the variables to satisfy a set of constraints.}
\end{marginnote}
how 
to determine the synaptic weights that let all input-output constraints
be satisfied simultaneously. 
%The network has a maximal {\it capacity}, typically $\alpha=M/N$, where $M$ is the 
%number of examples presented and $N$ their dimension.
Deep neural networks are able to extract high-level features from data and thus be notably efficient in the 
classification tasks. 
In the {\it random perceptron} the inputs and
associated outputs are taken to be uncorrelated random variables~\cite{Gardner88}. 
%(This is  not a {\it machine learning} problem,
%since even though all input-output relations could be correctly classified, it is impossible to generalise to new input-output patterns.)
If, moreover, the synaptic variables are  {\it spherically} constrained, the zero capacity 
corresponds to a satisfiability/unsatisfiability 
threshold which is analogous to the jamming transition of hard-spheres in very high dimensions.
The Schwinger-Dyson equations ruling the Langevin dynamics of a spherical random perceptron, 
with the same structure as 
Equations~\ref{eq:dyn-eqs-R}-\ref{eq:dyn-eqs-z}
have been established in~\cite{Agoritsas18}. The relation between the kernels $\Sigma$ and $D$ 
and the global correlation and linear response $C$ and $R$
are much more complicated than the simple algebraic ones of the $p$ body model. A complete analytic and 
numerical solution of this problem is not available yet but constitutes an interesting open problem.

The supervised learning dynamics of a deep neural network shares similarities but also differs from the relaxation of the 
spherical $p\geq 3$ body model~\cite{Baity-Jesi19}. 
\begin{marginnote}[]
\entry{Supervised learning}{uses a training set to teach the network to yield the desired output.}
\end{marginnote}
In an over-parametrized deep neural network,
in which the number parameters exceeds the size of the training dataset, 
three regimes were identified: an initial exploration of the energy/loss landscape, 
followed by a decrease of the loss in which the system displays aging dynamics, and 
a final almost stationary and diffusive regime in which the network  
reaches the bottom of the landscape. Barrier crossing
does not seem to play a role in the learning process while slow evolution along flat directions does.
Instead, in under parametrized networks the loss function approaches a non-vanishing value, 
there is aging, and the relaxation resembles strongly the one of the  spherical $p\geq 3$ model. 
%This transition is similar to the one found in an inference algorithm~\cite{Zdeborova16}.

A Langevin algorithm intended to sample the posterior probability measure for the spiked mixed matrix-tensor 
model, and its representation in terms of Schwinger-Dyson equations, was presented and 
studied in detail in~\cite{Sarao19,Sarao20}. This problem is very similar to the mixed $p+2$ model
discussed in Sec.~\ref{subsubsec:mixed}. 

 {\it Gradient descent} is typically used to train a network, but this requires the evaluation of the state 
of the neural weights on the full training set. 
\begin{marginnote}[]
\entry{Stochastic gradient descent}{Evaluates the gradient on a subset of the training set.}
\end{marginnote}
{\it Stochastic gradient descent}  approximates 
the gradient by evaluating it only on a small subset of the training set  (batch), 
changed at each step of the dynamics. Networks trained this way
have an excellent performance, and can even {\it generalize}, that is, classify correctly
 previously unseen data. 
 \begin{marginnote}[]
\entry{Generalization}{the ability to classify correctly
 previously unseen data.}
\end{marginnote}
 The choice of different batches introduces noise on the otherwise
 deterministic gradient descent. Understanding the nature of this noise is a hot topic.  
 This microscopic dynamics bear similarities with the ones 
of physical systems, where the stochastic dynamics of the degrees of 
freedom is dressed with local and independent forcing variables
 (strain deformation in driven systems and self-propulsion in active matter).
 %The trajectories of  stochastic gradient descent can be tracked analytically with DMFT. 
 Several stochastic gradient descent 
 algorithms which achieve a binary classification of Gaussian mixtures were 
 studied in this setting~\cite{Mignacco20,Mignacco22}.
 The $C$ and $R$ of the network weights can then be calculated~\cite{Mignacco22} and,
 more importantly, compared so as to extract a $T_{\rm eff}$ from the 
 deviations from the FDT~\cite{CuKuPe97}. 
 In the under-parametrized or un-satisfiable phase 
 where the network cannot achieve zero training error, the dynamics go to a stationary state
 with a finite $T_{\rm eff}$ with a non-monotonic dependence on the batch size. Instead, in the 
 over-parametrized or satisfiable phase the dynamics stop 
 at one solution with zero training error, and $T_{\rm eff}$ approaches zero asymptotically, 
 consistently with the idea that 
 stochastic gradient descent implements a self-annealing procedure. Reference~\cite{Mignacco22}
 discusses the behavior of $T_{\rm eff}$ in several improvements upon the simplest 
 stochastic gradient descent algorithm as well.

\section{QUANTUM DYNAMICS}
\label{sec:quantum}

Disorder and interactions  lead to the very rich phenomena discussed so far.
Adding quantum fluctuations amplifies the palette of intriguing phenomena even more so~\cite{Muller23}.

\subsection{Statics and dynamics in $p$ models}

Two parameters control the canonical equilibrium properties of quantum models with Hamiltonian 
$\hat {\mathcal H} = \hat K + \hat V$. One quantifies thermal fluctuations and 
it is the traditional temperature over the interaction strength $J$. The 
other one measures the strength of quantum fluctuations and can be organized  in the adimensional 
form $\Gamma=\hbar^2/(mJ)$. 

The $p$ body models have a static phase diagram with 
ordered and disordered phases. The order of 
the phase transition depends on the type of interactions considered. It is continuous for $p=2$, while it has a mixed
RFOT/truly first order transition for $p\geq 3$. The former applies close to the classical limit, while the latter close to the 
quantum critical point. These conclusions have been drawn from the 
replica analysis~\cite{CuGrSa00,CuGrSa01} which, on top of the usual order parameter in the form of an overlap matrix in 
replica space $Q_{ab}$, includes an imaginary time dependent one 
$Q_{aa}(\tau) = N^{-1} [\langle \sum_{i=1}^N {\rm T}  \hat x^a_i(\tau) \hat x^a_i(0)\rangle] = q_d(\tau)$. 
(The average represents $\langle \dots \rangle = {\rm Tr} (\hat \rho \dots)/{\rm Tr} \hat \rho $,
with $\hat \rho$ the equilibrium density operator at inverse temperature $\beta$, 
$a, b=1, \dots, n$ are replica indices, $\tau$ is the periodic imaginary time
and ${\rm T}$ the time ordering operator.)
This function is determined by the equation
\begin{equation}
G_0^{-1}(\tau) q_d(\tau) =\delta(\tau) +  \int_0^\tau d\tau' \, \Sigma(\tau-\tau') q_d(\tau')
\; , 
\label{eq:imaginary-time}
\end{equation}
with $G_0^{-1}(\tau) = m\partial_{\tau}^2 + z(\tau)$, $\Sigma = -J^2/(2\hbar) \,  q_d^{p-1}$, 
and $q_d(0)=q_d(\beta\hbar)$. 
Close to the quantum critical point $(T=0, \Gamma_{\rm Q})$ of the $p\geq 3$ model
some off-diagonal elements of the 
replica matrix,  $Q_{ab}=N^{-1} \sum_{i=1}^N [\langle {\rm T} \hat x^a_ix^b_i\rangle]$ with $a\neq b$, 
jump from zero to a finite value, and this implies that the ground states on the two sides of the transition are essentially 
unrelated~\cite{CuGrSa00,CuGrSa01}.

Also interesting is to consider the effect of the coupling to a 
an ensemble of quantum harmonic oscillators with spectral density $I(\omega)$.
It manifests in the addition of a bath induced term to the self-energy $\Sigma$ which
in Fourier space reads $\tilde \Gamma_R(\omega) \propto \omega^s$ with Ohmic ($s=1$), super ($s>1$) and sub ($s<1$) Ohmic
possibilities encoded in this compact form. Their effect is highly non-trivial as it can 
make the ordered phase more stable, and thus occupy a larger portion of the phase diagram~\cite{Cugliandolo02}.

The concept of a free-energy landscape can also be applied to a system with quantum fluctuations~\cite{BiCu01}.
Its properties, as $T/J$ and  $\Gamma$ are varied, can be studied
with similar techniques to the ones used to analyze the classical landscape, with the addition of 
the imaginary-time dependence of $q_d(\tau)$. In the $p\geq 3$ model
case, a complex landscape with large multiplicity of saddles of all kinds subsists under 
quantum fluctuations.

The relaxation of a  quantum mean-field disordered model in contact with a quantum bath 
is still governed by Schwinger-Dyson  equations like the ones in Equations~\ref{eq:dyn-eqs-R}-\ref{eq:dyn-eqs-z} although 
with some subtle differences~\cite{CuLo98,CuLo99}. Firstly, $G_0^{-1}(t_1) = m\partial_{t_1}^2 + z(t_1)$
is a second order differential operator.
Secondly, a quantum bath gives rise to retarded contributions to the self-energy and 
vertex $\Sigma$ and $D$ that replace the terms proportional to $\eta$ in Equations~\ref{eq:dyn-eqs-R}-\ref{eq:dyn-eqs-z}
(but do not depend on $C$ and $R$). They  read
The explicit coupling to a quantum bath adds to this equation two extra time-delayed terms
\begin{eqnarray}
\Gamma_R(t_1-t_2) &=& 
- \theta(t_1-t_2) \int_0^\infty d\omega \, I(\omega) \, \sin[\omega(t_1-t_2)]
%4\frac{m\eta }{\pi} \frac{d}{d(t-t'} \frac{\Lambda}{(1+[\Lambda(t-t')]^2} \theta(t-t')
\; , 
\\
\Gamma_C(t_1-t_2) &=&  \int_0^\infty d\omega \, I(\omega) \, \coth\left(\frac{1}{2} \beta\hbar\omega\right) \, \cos[\omega(t_1-t_2)]
\; . 
\end{eqnarray}
%$I(\omega)$ is the spectral density of the bath and their particular time-dependencies vary according to the type of bath considered.
Thirdly, the part of the kernels $\Sigma$ and $D$ due to the interactions are not 
as simple as the ones in Equations~\ref{eq:Sigma-D}. For the model with $p$-body interactions they read
\begin{equation}
\Sigma = - \frac{J^2 p}{\hbar} \, {\rm Im} \left(C-\frac{{\rm i} \hbar }{2} R \right)^{p-1} 
\!\!\!\!\! , \qquad\qquad 
D = \frac{J^2 p}{2} \, {\rm Re} \left(C-\frac{{\rm i} \hbar }{2} R \right)^{p-1} 
\!\!\!\!\! . 
\end{equation}
Equilibrium initial conditions at inverse temperature $\beta_0$ can also be imposed.
To do it, one needs to introduce correlations living on the imaginary time axis and, moreover, 
correlations that mix real and imaginary times. In the end, the terms proportional to $J_0/T_0$
in Equations~\ref{eq:dyn-eqs-R}-\ref{eq:dyn-eqs-z} are replaced by integral terms involving these 
functions, with the addition of equations that govern their real and imaginary time evolutions. In particular, 
an equation like~\ref{eq:imaginary-time} determines the evolution of $q_d(\tau)$ with the periodic 
boundary condition $q_d(\beta_0\hbar)=q_q(0)$ fixed by the initial temperature.
The full set of Schwinger-Dyson equations is considerably more complicated
than the one for the classical case~\cite{CuLoNe19}, 
which is recovered by taking $\hbar\to 0$.
Finally, let us mention that the quantum $p$-spin model can also be 
connected to the Mode-Coupling description of super-cooled quantum liquids~\cite{Rabani}.

The solution to the Schwinger-Dyson equations was first derived in a limit in which the coupling 
to the bath is taken to be so weak that it does not modify the location of the phase transitions. 
It shows similarities and differences with the 
classical ones~\cite{CuLo98,CuLo99}. First of all the second-order differential operator
induces oscillations, the frequency of which depend on the control parameters.
In consequence, after quenches from disordered initial conditions to parameters that are not too far from the ordered
phase or within it, the $C$ and $R$ at short time differences (high-frequency dynamics) 
depend on $\Gamma$. 
In particular, there are oscillations in $t_1-t_2$ that smoothly decay in amplitude as the 
plateau is approached. Instead, these functions at long time differences (low-frequency dynamics)
are monotonic and qualitatively identical to those of their classical counterparts.
For $p\geq 3$ models, the quenched dynamics in the ordered phase hovers over saddles in the free-energy landscape 
and slowly relaxes its energy, as in the weak ergodicity breaking scenario, with aging.  The quantum 
FD  relation is the equilibrium one on short time differences but takes the classical
form, with an effective temperature $T_{\rm eff}$ that exceeds the one of the bath (even
at $T = 0$). The generation of a $T_{\rm eff}>0$ ``decoheres'' the 
dynamics renders this regime classical. 
The dynamical phase transition  precedes the thermodynamic one also quantum mechanically.
The effect of the bath is highly non-trivial as it can 
enlarge the extent of the ordered phase in the phase diagram~\cite{Cugliandolo02}.
One can also tune the initial state to be within one of the valleys in
the free-energy landscape and show that the subsequent dynamics remains confined
to it (at least for times which are not exponentially large in system size)~\cite{CuLoNe19, Thomson20}. 
Because of the first order phase transition of the $p\geq 3$ systems, 
{\it quantum annealing} is not expected to be useful for them~\cite{Jorg08,Bapst13}
which are, however, the ones that one would  wish to optimize since they 
are related to the celebrated $K$-satisfiability optimization problem~\cite{Monasson97,Monasson99}.

\subsection{Many-body localization}

Quantum glassiness and many-body localization have been related to each other 
in recent years. However, they are important differences between the two 
phenomena.

Non-interacting quantum particles in one dimension under
a disordered potential Anderson localize and stop diffusing on the scale of the elastic
mean-free path. Many-body localization (MBL) is its interacting 
counterpart~\cite{Basko2006}.
It relies on the discreteness of non-interacting excitations and the rareness of resonant
interactions, which then fail to induce diffusion of energy. MBL can only occur in
isolation, with no coupling to the continuum of a bath which would reinstate transport
and ergodicity. In contrast, glassy phases are stable against a bath,
as they rely on the rugged structure of the free-energy landscape. Moreover, 
MBL phases are highly susceptible to local inclusions where disorder is weak, 
which inhibit it in high dimensions and may even destroy it completely (the fate of 
MBL in 1d is a subject of debate). 
In contrast, glasses, like standard symmetry broken
phases, are rendered more stable in high $d$.
In short, both quantum glassiness and MBL entail non-ergodicity and impede
full thermalization but they do so for fundamentally different reasons.

\subsection{The SYK model}

Kitaev proposed to use the $p=4$ potential where the operators $\hat x_i$ are replaced by Majorana fermions 
as a toy model for near-extremal black holes~\cite{Kitaev1}. 
The relevant correlation is 
$q_d(\tau) = - N^{-1} \sum_{i=1}^N [\langle {\rm T} \hat \psi_i(\tau) \hat \psi_i(0)\rangle ] $, 
and satisfies Equation~\ref{eq:imaginary-time} with 
$G_0^{-1} = d_\tau$ and $\Sigma = J^2 q_d^3$.
The main reason to claim
a connection with gravity in a nearly AdS$_2$ background is that this model is not glassy, but becomes approximately conformal 
after dropping time-derivatives, in the 
low-frequency limit, with a time-reparametrization symmetry which is broken to SL(2,R). 
A {\it sigma model} describing the cost of reparametrizations in terms of a {\it Schwarzian action} was 
constructed~\cite{Maldacena2}, similarly to what was done in~\cite{ChCu07,ChChCuReSe04} 
for the time-reparametrizations in glassy models.
The model shares 
also two thermodynamic properties of such black holes: it has 
non-zero entropy as $T\to 0$ (after $N\to\infty$) and a linear in $T$ specific heat. 
Kitaev's model is very close to the model proposed by Sachdev and Ye to describe
non-Fermi liquid behavior in condensed matter systems~\cite{Sachdev}, and was thus dubbed SYK.
A discussion of more connections between the SYK and glassy models with a RFOT 
can be found in~\cite{Facoetti19}.

The SYK model is a {\it maximally chaotic/perfect scrambler} 
meaning that the bound $\lambda_T \leq 2\pi T /\hbar$ on the Lyapunov exponent
$\lambda_T$  (a consequence of the 
FDT~\cite{Tsuji,Pappalardi}), defined from the exponential growth of
out-of-time-order correlations~\cite{Larkin89,Maldacena2}, is saturated.
This kind of chaos was then studied in the disordered and marginal glassy phases of the spherical $p$ models~\cite{Bera22,Correale23}.
Quantum fluctuations were found to make the disordered phase less and the glassy 
phase more chaotic. In the classical limit $\hbar\to 0$ a crossover from strong to weak
chaos, as marked by a maximum in $\lambda_T$, arises well above $T_d$, concomitant with the onset
of two-step slow relaxations. Finally, we comment that the experiment of setting 
two systems in contact, commonly performed in studies of disordered systems, see {\it e.g.}~\cite{FrPa97,CuKuPe97},
has, in this context, the interpretation of creating a wormhole between the two 
spaces.

\section{CONCLUSIONS}
\label{sec:Conclusions}

We discussed a number of classical systems, made up of a large number of degrees of freedom, 
with stochastic or Newtonian dynamics influenced by a very rough free-energy landscape, 
or complex quenched random forces. All this leads to uncommon phases
and even more unusual dynamics. Their quantum extensions 
are equally rich.

The models we dealt with may seem artificial for physical applications. Still, 
they provide a successful mean-field 
description of coarsening, glassiness, and jamming. Their  limitations to 
capture the full behavior of finite dimensional physical systems have been
identified and efforts are being made to overcome them. 
The same models are the actual problems one needs to understand 
in other fields of science, such as optimization, artificial intelligence, and ecology. The 
enormous amount of knowledge gathered by the statistical physics community 
can now be exploited in these other areas.

Some important and generic 
notions that came out of the analytic solution of 
the fully-connected models are 
{\it rugged free-energy landscapes}, 
{\it marginally stable regions}, 
{\it critical slowing down}, 
{\it weak and strong ergodicity breaking}, 
{\it weak long-term memory}, 
{\it aging}, 
{\it violations of the fluctuation-dissipation theorem},  
{\it effective temperatures}, 
{\it time reparametrization invariance},
{\it soft modes and dynamic fluctuations},
{\it Generalized Gibbs Ensembles}, 
{\it multiple fixed points},  
and {\it chaos}. 
They all appear in one way or another 
in the problems discussed here. Some of them were known beforehand but others not. Once uncovered in this 
idealized theoretical context, they were explored, and in some cases confirmed, numerically and experimentally 
in more realistic physical systems. 

Many open issues remain to be solved. The complete understanding of models with even more complex 
free-energy landscapes than the ones with monomial $p$ potentials is lacking. 
An efficient algorithm able to solve the generic self-consistency single-variable 
equations for the $d\to\infty$ problems, when no stationarity hypothesis can be used, is not available yet.
All analytic results were derived having taken the limit $N\to\infty$
from the start, and eventually considering the long-times asymptotic.
The sharp dynamic transitions thus predicted  cannot exist  for finite size systems or in
finite dimensions and should be replaced by a crossover. Having said so, 
the analytic description of this process, and the eventual approach to equilibrium 
appears as a formidable task and has remained elusive since the derivation of the 
$N\to\infty$ solution.  

\comments{
% Summary Points
\begin{summary}[SUMMARY POINTS]
\begin{enumerate}
\item Summary point 1. These should be full sentences.
\item Summary point 2. These should be full sentences.
\item Summary point 3. These should be full sentences.
\item Summary point 4. These should be full sentences.
\end{enumerate}
\end{summary}

% Future Issues
\begin{issues}[FUTURE ISSUES]
\begin{enumerate}
\item Future issue 1. These should be full sentences.
\item Future issue 2. These should be full sentences.
\item Future issue 3. These should be full sentences.
\item Future issue 4. These should be full sentences.
\end{enumerate}
\end{issues}
}

%Disclosure
\section*{DISCLOSURE STATEMENT}
The authors are not aware of any affiliations, memberships, funding, or financial holdings that
might be perceived as affecting the objectivity of this review. 

% Acknowledgements
\section*{ACKNOWLEDGMENTS}
The author thanks all her collaborators for exchanges and joint work performed over 
many years of exciting research.  She especially thank I. Agoritsas, A.  Altieri, T. Arnoulx de Pirey and F. Zamponi  for comments on the 
manuscript and A. Altieri and D. Barbier for help with the figures.
This research was supported in part by ANR-19-CE30-0014 and ANR-20-CE30-0031.

\bibliographystyle{ar-style4}
\bibliography{annual-reviews-LFC}

\end{document}